\documentclass[sigconf]{acmart}

\def\BibTeX{{\rm B\kern-.05em{\sc i\kern-.025em b}\kern-.08emT\kern-.1667em\lower.7ex\hbox{E}\kern-.125emX}}

\usepackage{listings} 
\usepackage{amsthm}

\usepackage{enumitem}
\usepackage{booktabs}
\usepackage{hyperref}

\usepackage{savesym}
\usepackage{multirow}
\usepackage{dcolumn}
\newlength{\thinline}
\newlength{\thickline}
\usepackage{colortbl}
\definecolor{mygray}{gray}{0.9}

\usepackage{amsmath}
\usepackage{subfigure}
\usepackage{graphicx}
\usepackage{multirow}
\usepackage{makecell}
\usepackage[english]{babel}
\usepackage{amsthm}

\usepackage{threeparttable}
\usepackage[utf8]{inputenc}
\usepackage{tcolorbox}

\newtheorem{Definition}{Definition}

\newcommand{\pwddict}[0]{\chi^{\ast}}

\begin{document} 

\fancyhead{}
\title{On the Account Security Risks Posed by Password Strength Meters}


\author{Ming Xu}
\affiliation{%
  \institution{Fudan University}
}
\affiliation{%
  \institution{National University of Singapore}
}
\email{xum18@fudan.edu.cn}

\author{Weili Han~\textsuperscript{*}}
\affiliation{%
  \institution{Fudan University}
}
\email{wlhan@fudan.edu.cn}

\author{Jitao Yu}
\affiliation{%
  \institution{Fudan University}
}
\email{21210240046@m.fudan.edu.cn}

\author{Jing Liu}
\affiliation{%
  \institution{UC Irvine}
}
\email{jing6@acm.org}

\author{Xinyi Zhang}
\affiliation{%
  \institution{Meta}
}
\email{xinyizhang@fb.com}

\author{Yun Lin}
\affiliation{%
  \institution{Shanghai Jiao Tong University}
}
\email{lin_yun@sjtu.edu.cn}

\author{Jin Song Dong}
\affiliation{%
  \institution{National University of Singapore}
}
\email{dcsdjs@nus.edu.sg}

\thanks{\textsuperscript{*} Corresponding author.} 



\begin{abstract} 

Password strength meters (PSMs) have been widely used by websites to gauge password strength, encouraging users to create stronger passwords. Popular data-driven PSMs, e.g., based on Markov, Probabilistic Context-free Grammar (PCFG) and neural networks, alarm strength based on a model learned from real passwords. Despite their proven effectiveness, the secure utility that arises from the leakage of trained passwords remains largely overlooked. To address this gap, we analyze 11 PSMs and find that 5 data-driven meters are vulnerable to membership inference attacks that expose their trained passwords, and seriously, 3 rule-based meters openly disclose their blocked passwords. We specifically design a PSM privacy leakage evaluation approach, and uncover that a series of general data-driven meters are vulnerable to leaking between $10^4$ to $10^5$ trained passwords, with the PCFG-based models being more vulnerable than other counterparts; furthermore, we aid in deriving insights that the inherent utility-privacy tradeoff is not as severe as previously thought. To further exploit the risks, we develop novel meter-aware attacks when a clever attacker can filter the used passwords during compromising accounts on websites using the meter, and experimentally show that attackers targeting websites that deployed the popular Zxcvbn meter can compromise an additional 5.84\% user accounts within 10 attempts, demonstrating the urgent need for privacy-preserving PSMs that protect the confidentiality of the meter's used passwords. Finally, we sketch some counter-measures to mitigate these threats.

\end{abstract}

\maketitle







\section{Introduction}   
Textual passwords remain the primary authentication method for websites across various domains, including news~\cite{Golla:18ccspasswordmeter,DBLP:conf/sp/MunyendoAA23:kenya}, email~\cite{DBLP:journals/scn/XuH19}, or financial services~\cite{DBLP:conf/sp/HuangAFHJ15, DBLP:journals/tissec/CarnavaletM15}. To combat weak password creation, almost every website utilizes a password strength meter (PSM) to provide strength feedback, encouraging users to create stronger passwords. 

\begin{figure}[htbp]
\centering
    \subfigure[Rule-based meters]{\includegraphics[width=0.16\textwidth]{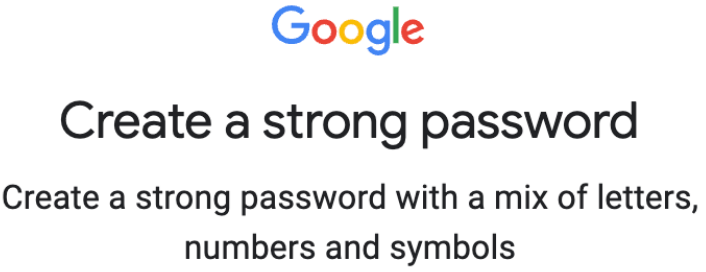}} 
    \subfigure[CKL\_PSM: data-driven meters]
    {\includegraphics[width=0.26\textwidth]{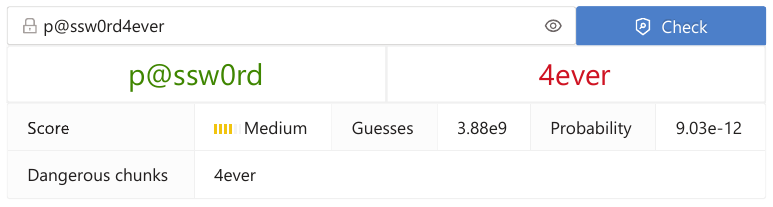}} 
    \caption{Examples of rule-based and probabilistic meters.\label{fig:metercases}} 
\end{figure}

PSMs have been widely recognized as effective in numerous studies~\cite{meter:userful,Weir:test,Golla:18ccspasswordmeter,Wang23,DBLP:journals/tifs/GalballyCS17}, and are extensively adopted by industry-leading platforms such as Google and Dropbox. Typically, PSMs can be divided into rule-based meters~\cite{(SP)80063:Intro, keepsm, DBLP:conf/uss/KomanduriSCHS14:telepathwords,zxcvbn,DBLP:conf/chi/UrAABCCCDNHJM17:explainable-meter} and data-driven meters~\cite{DBLP:conf/ndss/CastellucciaDP12:adaptive-markov,DBLP:conf/dsn/WangHCW16:fuzzypsm,DBLP:conf/ccs/XuWYZZH21:chunk} (shown in Figure~\ref{fig:metercases}). 
Rule-based meters evaluate strength based on predefined rules like length, character classes, blocklist check, or the identification of weak patterns (e.g., keyboards). It is worth noting that the rule of blocklist check is more recommended by recent studies~\cite{DBLP:conf/ccs/TanCCS20,DBLP:conf/chi/UrAABCCCDNHJM17:explainable-meter,blocklist-ndss,zxcvbn} due to the balance of usability and security. 
Zxcvbn~\cite{zxcvbn}, a commercial rule-based meter adopted by Dropbox since 2016, evaluates strength by detecting blocked passwords and weak patterns. Rule-based meters are known for simplicity, yet may sometimes yield inaccurate evaluation for human-created passwords: they may label ``p@ssw0rd'' as a strong password while deeming a random string like ``dasglkew'' as weak based on character classes.
Researchers later introduced data-driven PSMs that leverage probabilistic password models trained on real-world passwords to effectively capture users' adaptive password creation patterns. By estimating the number of attempts required to guess a password---where a higher number of guesses indicates stronger security---these models provide both adaptability and accuracy in evaluating human-generated passwords, capable of tailoring to the specific requirements of a user community. For instance, an organization could train community-specific data-driven meters using the internal, nonpublic password data to provide more accurate feedback and enhance the security of its internal password ecosystem.
Data-driven models typically include statistical approaches like Markov~\cite{DBLP:conf/ndss/CastellucciaDP12:adaptive-markov} and Probabilistic Context-Free Grammars (PCFG)~\cite{DBLP:conf/dsn/WangHCW16:fuzzypsm,DBLP:conf/ccs/XuWYZZH21:chunk}, as well as neural networks~\cite{William:LSTM, DBLP:conf/esorics/PasquiniAB20:CPGmeter}.

Despite their widespread adoption, both rule-based and data-driven PSMs pose significant security risks, for example, they may expose the passwords used for training or blocking to malicious attackers, leading the search space for those account employing the meters narrowed down. 
Several real-world PSMs prioritize accuracy over privacy by leveraging non-public data. For instance, Kaspersky Password Manager~\cite{Kaspersky-password-manager} and RoboForm~\cite{roboform} incorporate private data from the Have I Been Pwned (HIBP) service into their blocking mechanisms. However, HIBP contains sensitive breached credentials including data provided by the FBI~\cite{hunt2023genesis:FBI}, leading the potential leakage of private passwords to misuse.  
To the best of our knowledge, no comprehensive investigation has been conducted into the potential damage caused by privacy vulnerabilities.

In this paper, we introduce a new threat model in which an adversary can download a data-driven model deployed on the client side, make repeated queries to bypass rate-limiting mechanisms, and infer the training status of specific passwords, thereby facilitating password theft. This attack exploits the tendency of data-driven models to \emph{over-learn}, assigning higher probabilities to passwords included in the training data. 
Since PSMs are typically deployed client-side (ensuring the server never receives the user's input password), this threat model is both realistic and widely applicable. Drawing on terminology from prior work in machine-learning domains, we classify these attacks as \emph{membership inference attacks}.
Second, we emphasize that the leakage of trained or blocked (referred to as ``used'') passwords specific to a meter significantly heightens risks to user accounts relying on the meter. A motivated attacker could exploit these leaked passwords to refine their cracking strategies—for example, by excluding used passwords during candidate password generation—thereby increasing their success rate.
These security concerns may impede the widespread adoption of data-driven meters on large-scale websites. Moreover, without a systematic evaluation methodology, it is challenging to fairly compare PSMs or establish a standard for rating their privacy leakage.
To summarize, we focus on the following research questions:
\emph{RQ1: How effectively do different data-driven meters resist membership inference and password-stealing attacks?}
\emph{RQ2: What are the security implications of exposing used passwords specific to the meter?}

We conduct an empirical investigation on 5 data-driven meters including AdaptivePSM~\cite{DBLP:conf/ndss/CastellucciaDP12:adaptive-markov} based on Markov models, FuzzyPSM~\cite{DBLP:conf/dsn/WangHCW16:fuzzypsm} based on PCFG models, FLA PSM~\cite{William:LSTM} based on neural-network models, IPPSM~\cite{DBLP:conf/esorics/PasquiniAB20:CPGmeter} based on neural-network models and CKL\_PSM~\cite{DBLP:conf/ccs/XuWYZZH21:chunk} based on optimized PCFG models, and 3 rule-based meters including KeePSM~\cite{keepsm}, Zxcvbn~\cite{zxcvbn} and CUPS PSM~\cite{DBLP:conf/chi/UrAABCCCDNHJM17:explainable-meter}. 
The first challenge lies in the customized design of effective membership inference attacks to PSMs. We specifically analyze the characteristic of data-driven models, and propose three meter-specific approaches: the probability-threshold-selection approach that settles down the probability threshold based a shadow model (i.e., the same model trained with owned data to mimic the target models' behavior), the binary classifiers trained with features of a shadow model, and the straightforward application of the Salems' method~\cite{DBLP:conf/ndss/Salem0HBF019ML-leaks} that picks the passwords ranking the top k\% as the member passwords without a shadow model. 
We compare that the probability-threshold-selection method and binary classifiers can generally outperform the Salems' method, showcasing the effectiveness of the shadow models in this task. 
Besides, our results show that neural-network-based meters are more robust to membership inference attacks, Markov-based meters are the second, while PCFG-based meters seriously suffer from the attacks. 
To launch the theft of trained passwords, we employ Generative Adversarial Network (GAN) and dynamic GAN techniques to generate additional passwords, to closely resemble the trained ones and steal more trained passwords. 
Our empirical and theoretical analyses reveal that approximately $10^4$ \textasciitilde $10^5$ trained passwords can be stolen with high confidence across 5 data-driven meters.

Secondly, the used passwords of a PSM are case-sensitive that potentially cause several security ramifications, yet empirically validating these consequences is challenging due to their diverse nature. 
We empirically show that when a motivated attacker removes the used passwords of a meter, they can significantly increase account compromise rates for users employing that meter. 
For instance, with Zxcvbn's used passwords, an attacker can compromise an additional 5.84\% of accounts within just 10 attempts on websites deploying Zxcvbn, showcasing the risks of leaking meter's used passwords. 
This highlights the need to maintain confidentiality for passwords used within a meter. Finally, we outline counter-measures to address these risks for both rule-based and data-driven meters.


\noindent\textbf{Contributions.} We summarize our main contributions as follows:
\begin{itemize}
   \item \textbf{Used Password Inference.} 
   We propose an evaluation framework to assess how well different data-driven meters resist inference and stealing attacks, comparing their vulnerability to password leakage.  
   \item \textbf{Used Password Exploit.} We highlight a novel kind of threat model outlining that an motivated attacker can leverage the used passwords to increase risks for accounts tied to a PSM. We empirically simulate that the leakage of used passwords specific to a meter can cause the additional account compromise on websites utilizing that meter. 
    \item \textbf{Insights and Mitigation.} 
    We find that even non-easier passwords remain vulnerable to inference attacks; the utility and privacy in meters are not inherently contradictory. Finally, we propose countermeasures to address and mitigate these security threats.  
\end{itemize}



\noindent We aspire to raise awareness among several stakeholders when deploying these meters in real-world websites. 
Our attacks highlight broader classes of design malpractices
found in PSMs. We identify these higher-level issues, and outline a series of takeaways for application designers.
To foster further research in this field, we will make our code and models publicly available in~\footnote{\url{https://anonymous.4open.science/r/MIA4PSMs-0D55/README.md}}.


\section{Background and Related Works~\label{sec:background}}



\begin{table*}[htbp] 
\setlength{\abovecaptionskip}{0pt}
\setlength{\belowcaptionskip}{0pt}
\caption{A survey of PSMs through a scoping review of academic literature focusing on rule-based and data-driven one. Quantization are several ratings of strength, e.g., Q3=[Weak, Medium, Strong]. Type: C=Client, S=Server. Pwd prob represents the estimated probability by the meters.}  
\label{tab:surver-PSM}
\renewcommand\tabcolsep{5.6pt}
\footnotesize
\begin{tabular}{c|c|c|c|c|c|c|c|c|c|c}
\toprule
\rowcolor{gray!45} Type & Meters   & \begin{tabular}[c]{@{}l@{}}Published\\ Year\end{tabular} & 
Deployment  & \begin{tabular}[c]{@{}l@{}}Blocklist\\ Leakage\end{tabular}  & Adaptive &\begin{tabular}[c]{@{}l@{}}Feedback\\ Form\end{tabular} & \begin{tabular}[c]{@{}l@{}}Strength\\ Evaluation\end{tabular}
 & Type & Approaches & Quantization \\
\midrule  
Rule & NIST-PSM~\cite{(SP)80063:Digital} & 2006 & Industry  &  $\times$  & $\times$   & Entropy  &  \checkmark  & C & Rules & Q3   \\
Rule & Telepathword~\cite{DBLP:conf/uss/KomanduriSCHS14:telepathwords}  &  2013  &  Microsoft   &  $\times$         & $\times$      & Next character & $\times$ & S & Trie & N/A \\ 
Rule & KeePSM~\cite{keepsm}           & 2015 & KeePass   & 10,183 & $\times$   & Entropy & \checkmark  & C & Patterns & Q5  \\  
Rule &Zxcvbn~\cite{zxcvbn} & 2016 & Dropbox  & 47,023   & $\times$     & Guess Number & \checkmark    & C & Patterns & Q5 \\
Rule & LPSE~\cite{DBLP:journals/compsec/GuoZ18:LPSE}              & 2018 & Academia    &  $\times$  & $\times$   & Similarity & \checkmark  & C & Cosine similarity & Q3   \\ 
\midrule
Data-driven & AdaptivePSM~\cite{DBLP:conf/ndss/CastellucciaDP12:adaptive-markov}   & 2012 &   Academia   &   $\times$     & \checkmark    & Pwd prob & \checkmark       & C  & 4-gram  & Q2  \\
Data-driven & FuzzyPSM~\cite{DBLP:conf/dsn/WangHCW16:fuzzypsm}          & 2016 & Academia    &  $\times$          & \checkmark   & Pwd prob  & $\times$      & C & FuzzyPCFG & N/A   \\
Data-driven &FLA PSM~\cite{William:LSTM}      & 2016 & Academia                                                                   &  $\times$      & \checkmark    & Pwd prob &  \checkmark & C/S & LSTM & Q3 \\
Data-driven &IPPSM~\cite{DBLP:conf/esorics/PasquiniAB20:CPGmeter}             & 2020 & Academia   &  $\times$   & \checkmark  & Pwd prob & $\times$  & C & Auto Encoders & N/A  \\ 
Data-driven & CKL\_PSM~\cite{DBLP:conf/ccs/XuWYZZH21:chunk}         & 2021 & Academia &  $\times$  & \checkmark  & Pwd prob & \checkmark & C/S & CKL\_PCFG & Q3 \\ 
\midrule
Combined & CUPS PSM~\cite{DBLP:conf/chi/UrAABCCCDNHJM17:explainable-meter} & 2017 & Academia  &  87,144 &  \checkmark      & Pass or not & \checkmark      & C & Rules and LSTM & Q2 \\

\bottomrule
\end{tabular}
\end{table*}

\subsection{Password Strength Meters}  
Password Strength Meters (PSMs) can date back to 1990s~\cite{DBLP:journals/tissec/BergadanoCR98:98PSM, DBLP:journals/compsec/BishopK95:95PSM} when they firstly check user-chosen passwords against a dictionary of weak passwords. Yan~\cite{DBLP:conf/nspw/Yan01:firstrulemeter} illustrated that such simple checkers can miss other weak passwords, and
proposed \textbf{rule-based meters} based on heuristic rules, such as password length, character types, a blocklist check, or identification of weak patterns like keyboard or repeat patterns.
Rule-based meters~\cite{DBLP:conf/ccs/Alroomi023:measure-password}, widely deployed in real-world websites such as Google or Dropbox, are renowned for their simplicity and efficiency. However, heuristic rules~\cite{WeirPCFG,kelley:guess} are unsuitable in measuring the strength of human-chosen passwords due to common tricks (i.e., users always simply change the ``password'' to ``p@ssw0rd''). 

Academic researchers recommended \textbf{data-driven password strength meters}. 
A data-driven model typically generates password candidates in descending order of probabilities, and the rank of a given password in the candidates represents the number of attempts required to guess that password, commonly referred to as the ``guess number''.
Common password models include Markov models~\cite{NarayananMarkov}, Probabilistic Context-free Grammars (PCFG)~\cite{WeirPCFG}, and neural-network-based models~\cite{William:LSTM}. Narayanan et al.~\cite{NarayananMarkov} first applied Markov models to password guessing in 2005. For a n-order Markov model, the probability of a character depends on the previous $n-1$ characters. 
On the other hand, in 2009, Weir~\cite{WeirPCFG} proposed PCFG models in password guessing.
PCFG computes probabilities using templates, where letter strings and digit strings are denoted as ``$L_n$'' and ``$D_n$'' respectively (where n indicates length). 
In 2016, Melicher et al.~\cite{William:LSTM} proposed Long Short Term Memory (LSTM) to build password models~\cite{William:LSTM} were proposed to model password guessing, which also referred to as FLA models due to the fast, lean and accurate characteristic. 
We illustrate the details in the following.

\noindent\textbf{Rule-based meters.} 
In 2006, National Institute of Standards and Technology (NIST) introduced NIST-PSM~\cite{(SP)80063:Intro} that employed ad-hoc rules to evaluate strength. 
Later in 2017, NIST SP800-63B~\cite{(SP)80063:Digital} recommended the length requirement and blocklist check to create secure passwords. Many studies~\cite{DBLP:conf/ccs/TanCCS20,DBLP:conf/chi/UrAABCCCDNHJM17:explainable-meter,blocklist-ndss} also recommended the blocklist check to balance security and usability.
KeePSM~\cite{keepsm} is a built-in password strength measurement based on heuristic rules inherent in KeePass, which is an open-source password manager that manages the user's passwords across websites.   
Zxcvbn~\cite{zxcvbn}, employed by the Dropbox company since 2016, treats a password as a sequence of tokens and evaluates every tokens by weak password blocklists and weak patterns like \emph{reversed} (e.g., 123321), \emph{repeat} (e.g., 123123), \emph{keyboard} (e.g., qwerty), or \emph{leet} (e.g., p@ssw0rd) to determine strength. 
In 2017, Ur et al~\cite{DBLP:conf/chi/UrAABCCCDNHJM17:explainable-meter} proposed a combined data-driven password meter (termed as CUPS PSM due to the deployment of CUPS demo website~\footnote{\url{https://cups.cs.cmu.edu/meter2/}}), based on 21 hybrid heuristic rules and the data-driven FLA model. The original implementation (v1.0) of this meter has been extended to include additional support for minimum-strength and blocklist requirements (v2.0)~\cite{DBLP:conf/ccs/TanCCS20}.
Besides, several rule-based meters have extended beyond traditional rules. For example, Telepathword~\cite{DBLP:conf/uss/KomanduriSCHS14:telepathwords}, implemented by Microsoft in 2013, predicts the next character of an input password, to discourage the expected character without strength evaluation. 
In 2018, LPSE~\cite{DBLP:journals/compsec/GuoZ18:LPSE} were proposed to use the cosine distance with weak passwords to evaluate strength.

\noindent\textbf{Data-driven meters.} Data-driven meters can be formalized as a learned function $f : \pwddict \to [0, 1]$, where $\chi^{\ast}$ is the set of all possible passwords. 
When trained on a leaked breach dataset $D_{train}$ (where $D_{train} \subseteq \pwddict$), a data-driven meter takes a password input $x$ ($x \in \pwddict$) and outputs an estimated probability denoted as $f(x)$, where $f(x) \in [0,1]$.   
A normalized PPSM adheres to the condition that the sum of estimated probabilities for all passwords in $\pwddict$ equals 1: $\sum_{x \in \pwddict} f(x)=1$. 
Generally, a password model learns statistical patterns like the likelihood of certain characters or combinations appearing together from a dataset of real-world passwords, enabling to generate new passwords that mimic these patterns. 
Let $C_{G}(f)$ denote the top $G$ candidates generated by the data-driven meter $f$, defined as follows:

\begin{Definition}\nonumber~\label{Definition2}
$C_{G}(f) = { x_1, x_2, \dots , x_G }$ with $f(x_1) \ge f(x_2) \ge \dots \ge f(x_G)$.
\end{Definition}

\noindent where $x_i \in \pwddict$. We denote $R_f(x_i)$ as the rank of the password $x_i$ in $C_{G}(f)$. Thus, $R_{f}(x_i)=i$.

The guess number $R_f(x_i)$ can usually be calculated by generating password candidates from $f$. This process is computationally intensive and requires substantial storage. 
Dell'Amico et al.~\cite{DellAmico:Monter_Carlo} proposed the Monte Carlo simulation methods to map the estimated probability to the guess number, with a prerequisite that $f$ must be normalized. 
In addition, the list model~\cite{DBLP:conf/sp/WangHoneyword2022, Wang23} is formally defined as: 
    $P(x)= \left\{\begin{matrix} \frac{Count(x)}{|D_{train}|},\quad &x \in D_{train} \\ 0, \quad &x \notin  D_{train}
      \end{matrix}\right.
          $
    where Count(x) means the occurrence number of the password x.  
    The list model is usually not deployed in a website for the following two reasons: 1) the list model cannot generalize to evaluate the strength of passwords outside of $D_{train}$; 2) The list model poses security risks by leaking the entire sets $D_{train}$. 
We then illustrate several types of data-driven meters below.

\begin{itemize}[fullwidth,itemindent=0em]
     
    \item\textbf{Markov-based meters:} 
    In 2012, Castelluccia et al.~\cite{DBLP:conf/ndss/CastellucciaDP12:adaptive-markov} first built a meter based on an adaptive Markov (termed as AdaptivePSM) with probability smoothing techniques. They randomly added noise to the statistical frequency, i.e., the statistical frequency is increased by one with a random probability $\gamma$.

    \item \textbf{PCFG-based meters:} 
    In 2016, Wang et al~\cite{DBLP:conf/dsn/WangHCW16:fuzzypsm} proposed FuzzyPSM based on FuzzyPCFG. FuzzyPSM is known for its high accuracy of strength evaluation.
    FuzzyPSM applies a base dictionary to learn the basic words $B$ and a training dictionary to model the fuzzy transformation grammars. 
    For example, the probability of the password ``w0rd123'' is calculated as $P(B_4B_3)\times P(B_4 \rightarrow word)\times P(B_3 \to 123) \times P(o \to 0)$, where the probability of transformation rules like $o \rightarrow 0$ is defined by their pre-defined statistics. 
    
    In 2021, Xu et al.~\cite{DBLP:conf/ccs/XuWYZZH21:chunk} proposed CKL\_PSM based on the chunk-level PCFG. CKL\_PSM uses the BPE (Byte Pair Encoding) algorithm to divide a password into several chunks, and then learn the chunk-segmented PCFG grammar. For example, ``w0rd123'' is first segmented as ``w0rd 123'', where probability is calculated as $P(DM_4D_3) \times P(DM_4 \rightarrow w0rd)\times P(D_3 \rightarrow 123)$, where $DM$ refers to a \underline{d}ouble \underline{m}ixed template. 

    \item \textbf{Neural-network-based Meters:}
    In 2016, William et al.~\cite{William:LSTM} proposed to build a FLA meter (termed as FLA PSM) based on the LSTM model. FLA PSM predicts the probabilities from left to right, for example, $P(c_1c_2 \dots c_l)=\prod_{i=1}^{l}P(c_i|c_{\mathit{start}},\dots,c_{i-1})$.
    
    In 2021, Pasquini et al.~\cite{DBLP:conf/esorics/PasquiniAB20:CPGmeter} proposed an Interpretable-Probabilistic-Password-Strength-Meter (termed as IPPSM) based on an auto-encoder architecture with CNN blocks. Formally, IPPSM calculates the probability of each character based on its left and right context as follows:  $$ P(c_i)= \begin{cases} P(x_i|c_2,c_3,\dots,c_l),\quad & i=1 \\ 
    P(x_i|c_1,c_2,\dots,c_{l-1}),\quad & i=l \\ 
    P(x_i|c_1,c_2,\dots,c_{i-1},c_{i+1},\dots,c_l),\quad & 1 < i < l 
    \end{cases}$$
    \end{itemize} 
    

Since 2012, researchers have explored data-driven password strength meters, beginning with AdaptivePSM~\cite{DBLP:conf/ndss/CastellucciaDP12:adaptive-markov}, to evaluate human-created passwords in contrast to rule-based meters. AdaptivePSM employed a noise mechanism to formally protect the n-gram database from leakage, addressing the inherent sensitivity of substring patterns in n-gram models. However, its large storage requirements made it impractical for client-side deployment. To address this, researchers later introduced compression techniques, leading to models like FLA in 2016, which enabled real-time strength inference. Neural-network-based models like FLA do not expose sensitive substrings through their encoded parameters, making them more suitable for industrial deployment.
However, the deployment challenge is exacerbated by a range of privacy concerns: while plaintext leakage of underlying password statistics such as n-grams is removed in FLA PSM, the individual privacy that an attacker downloads the model and launch an offline MIA attack to steal the trained passwords remains insufficiently explored. Especially, in practice, industry adoption often involves customized data-driven meters trained on proprietary datasets to capture community-specific password patterns~\cite{Li14,Wang23,DBLP:journals/corr/abs-2301-07628-SP-2024-Pasquini}, making it more important to investigate the privacy risks.  The privacy implications of such deployments remain insufficiently examined, leaving a critical gap that impedes widespread adoption. 


\subsection{Security Risks.}

\noindent\textbf{Data leakage risks in rule-based meters.} 
As shown in Table~\ref{tab:surver-PSM}, KeePSM, Zxcvbn and CUPS PSM
openly expose their specifically blocked password dictionary to approximately $10^4$ size on the client-side, demonstrating potential password leakage risks.
While LPSE~\cite{DBLP:journals/compsec/GuoZ18:LPSE} also leverages a weak password dictionary, it converts blocked passwords into numerical vectors. While such measures may provide some level of confidentiality for used passwords, there remain risks associated with potential attacks when attackers break the vector databases. 

\noindent\textbf{Membership inference attacks in data-driven meters.} Membership inference attacks (MIAs) aim to ascertain whether a record was utilized during the model training, and are received widely attention in machine learning models~\cite{DBLP:conf/sp/Shokri17,DBLP:conf/ndss/Salem0HBF019ML-leaks,DBLP:conf/icml/Choquette-ChooT21:label-MIA, DBLP:conf/ndss/Salem0HBF019ML-leaks}. Formally, the attack model of MIAs can be descried as $f(x,\theta) \rightarrow \{in, out\}$, where 
$f(x, \theta)$ refers to the inference outcome of the given record $x$ on the model $\theta$, $in$ and $out$ denote the member and non-member status. 
In 2017, Shokri et al~\cite{DBLP:conf/sp/Shokri17} first proposed a binary-classifier to launch MIAs against machine learning models. Later on, Salem et al~\cite{DBLP:conf/ndss/Salem0HBF019ML-leaks} weakened the original assumption of Shokri's models and proposed a lightweight attack model. It is intuitive that data-driven meters suffer from the MIAs,
unfortunately, unlike machine learning models that provide prediction vectors with probabilities for each output class, data-driven password models typically return only a single probability, limiting the direct applicability of MIAs in machine-learning methods to data-driven password models. While there are techniques for conducting MIAs against machine learning models with a single probability class~\cite{DBLP:conf/icml/Choquette-ChooT21:label-MIA}, they often rely on changes in the loss function during model training. Most statistic password models do not have loss functions. As a result, methods for MIAs used in machine learning models cannot be directly applied to data-driven meters.
To resist MIAs, Pasquini et al.~\cite{DBLP:journals/corr/abs-2301-07628-SP-2024-Pasquini} introduced a universal guessing model and specifically developed an extension with a differential privacy mechanism in 2024, targeting to the neural-network-based models. 
There remains a significant gap in building more robust data-driven meters including statistical approaches.

\subsection{Datasets}
Since 2009, many websites have suffered from password leaks, with these leaked datasets serving as the corpus for password researches. We use these publicly available datasets, which have been widely used in previous studies, to investigate the security issues in this paper.   
We summarize our used datasets in Table~\ref{tab:datasets}.
\texttt{178}~\cite{178} is a dataset leaked from a Chinese game website, while \texttt{Rockyou}~\cite{Rockyou} comes from a English game website. \texttt{XATO}~\cite{xato} is a comprehensive dataset which comprises data from multiple breaches of English websites. 
\texttt{Cit0day}~\cite{cityday} is a recent leaked dataset mostly from English users. We preprocess these datasets by removing the non-ASCII and abnormally long passwords as done by~\cite{DBLP:journals/scn/XuH19, DBLP:conf/uss/PasquiniCAB21:dynamic}.

To investigate the effect of the used passwords in compromising accounts on the websites employing the meter, we simulate a targeted guessing scenario using two datasets of \texttt{4iQ}~\cite{4iQ} and \texttt{Collection\#1}~\cite{collection1}, which contain Emails. In this context, we identify an Email as a unique account. We pre-process these raw datasets by merging passwords associated with the same Email~\cite{DBLP:conf/sp/PalD0R19:similarity, xu-real-world-guessing}, to identify all those passwords belonging to the same user.
As detailed in Table~\ref{tab:datasets}, after merging the \texttt{4iQ} dataset, it contains $147,284,401$ unique accounts (Emails) with a total of $373,820,141$ passwords.
This suggests that, on average, each user possesses approximately 2.538 passwords.

\noindent\textbf{Ethical concerns.} Our work only presents the statistical information for the requirement of ethical practice. While we use real-world datasets that include Emails, we do not identify the exact user associated with the leaked passwords. Instead, we focus on the whole feature collection of many user's passwords in a breached dataset.
Although the leak is publicly available on Internet, we do not want to publicize it and process the datasets by a computer not connected to internet.
Since our datasets are all publicly available from various sources over the Internet, the results in this work are reproducible.


\begin{table}[!bht]
\setlength{\abovecaptionskip}{0pt}
\setlength{\belowcaptionskip}{0pt}
\footnotesize
\centering
\caption{\label{tab:datasets}Basic information about the password datasets used in this paper. Pwds refer to passwords.} 
\renewcommand\tabcolsep{8.7pt}
\begin{tabular}{ccccc}
\toprule[\thickline]
\rowcolor{gray!45} Language & Dataset & Year & Valid Pwds & \begin{tabular}[c]{@{}l@{}}Unique Pwds\\ (Accounts)\end{tabular}  \\
\cmidrule{2-5}
Chinese & \texttt{178} & 2009    & 9,071,979           &  3,461,974  \\
 \midrule[\thinline] 
\multirow{5}{*}{English}  & \texttt{Rockyou} &  2009 &  32,582,532 & 14,270,373  \\ 
 & \texttt{XATO}  & 2015  &  9,991,998 & 5,186,444    \\ 
\cmidrule{2-5}
& \texttt{4iQ} & 2017 &  373,820,141 & 147,284,401  \\  
 & \texttt{Collection\#1} & 2019 &  365,336,365 & 109,191,685  \\ 
 \cmidrule{2-5}
 & \texttt{Cit0day} &  2020 & 86,835,796  & 40,589,949  \\ 
\bottomrule[\thickline]
\end{tabular}
\end{table}

\section{Membership Inference and Password Stealing Attacks~\label{sec:meter-attacks}} 


\subsection{Threat Model} 
The adversary can be potentially any legitimate user of a website, and is able to download the password model behind the meter. Therefore, the adversary can freely attempt random passwords to the downloaded offline model with unlimited number of times, rendering rate-limiting countermeasures ineffective. This threat model is realistic as in practice, because existing meters are always deployed in client-side.  

\noindent\textbf{Adversary's knowledge and goals.}
Adversaries typically have black-box access to a data-driven meter, enabling them to input a password and receive an output probability, where they can obtain the estimated probability either directly or by briefly analyzing the model. 
We assume that adversaries have access to one leaked dataset (i.e., \texttt{178} or \texttt{XATO}) as their prior attack knowledge and are aware of the model's architecture, like whether it's Markov-based or PCFG-based.

The distinction between MIAs and stealing attacks lies in their objectives: membership inference attacks aim to distinguish whether a password is trained (member) or not (non-member), while stealing attacks focus on generating passwords that closely resemble trained ones, to construct the training database. 
On the basis of MIAs, adversaries seek to infer more trained passwords from the meter, going beyond the attack knowledge of the owned dataset. 
To this end, the adversary can employ modern deep learning techniques to augment datasets that are more likely to belong to the set of trained passwords. The adversary leverages the additionally generated passwords to query the target meter, and distinguishes the member passwords via MIAs' model to form the training database.

\subsection{Motivating Principle}  
Here, we illustrate our motivating principles regarding the meter's over-learning phenomenon. 

\noindent\textbf{Over-learning phenomenon.}~\label{sec:principle-MIA}  
Generally, we find that, in the case of data-driven models, trained passwords tend to be estimated with especially high probabilities. This is because that the fitting degree of the training passwords is too high.  
In the field of machine learning, researchers have described such undesirable behavior as ``over-fitting'', and already analyzed the connection between over-fitting and privacy risks~\cite{2018/overfitting/machine/learning}. However, limited password research discussed the privacy implications of such undesirable behaviors. In this paper, we term it ``over-learning'' we posit that it's crucial to differentiate this concept in the context of password modeling.

Over-learning in password context was first noticed in ~\cite{Ma:}, where it was observed that higher-order Markov models exhibit more pronounced degree of fitting to trained passwords. 
The issue of over-learning can compromise both the accuracy and security requirements of a model. In terms of accuracy, an excessively high degree of fitting training data can restrict the model's capacity to evaluate the strength of passwords not present in the training sets. Accuracy typically captures the general password distribution ($\pwddict$), but over-learning indicates that the meter only learns from the distribution within the training sets ($D_{\mathit{train}}$).
In terms of security, over-learning can result in significant differences between trained and non-trained passwords, making them easily distinguishable based on their probabilities.

\begin{figure*}[!htb]
    \centering
    \subfigcapskip=-6pt
        \subfigure[\texttt{3-order Markov (4-gram)}]{\includegraphics[width=0.3\textwidth]{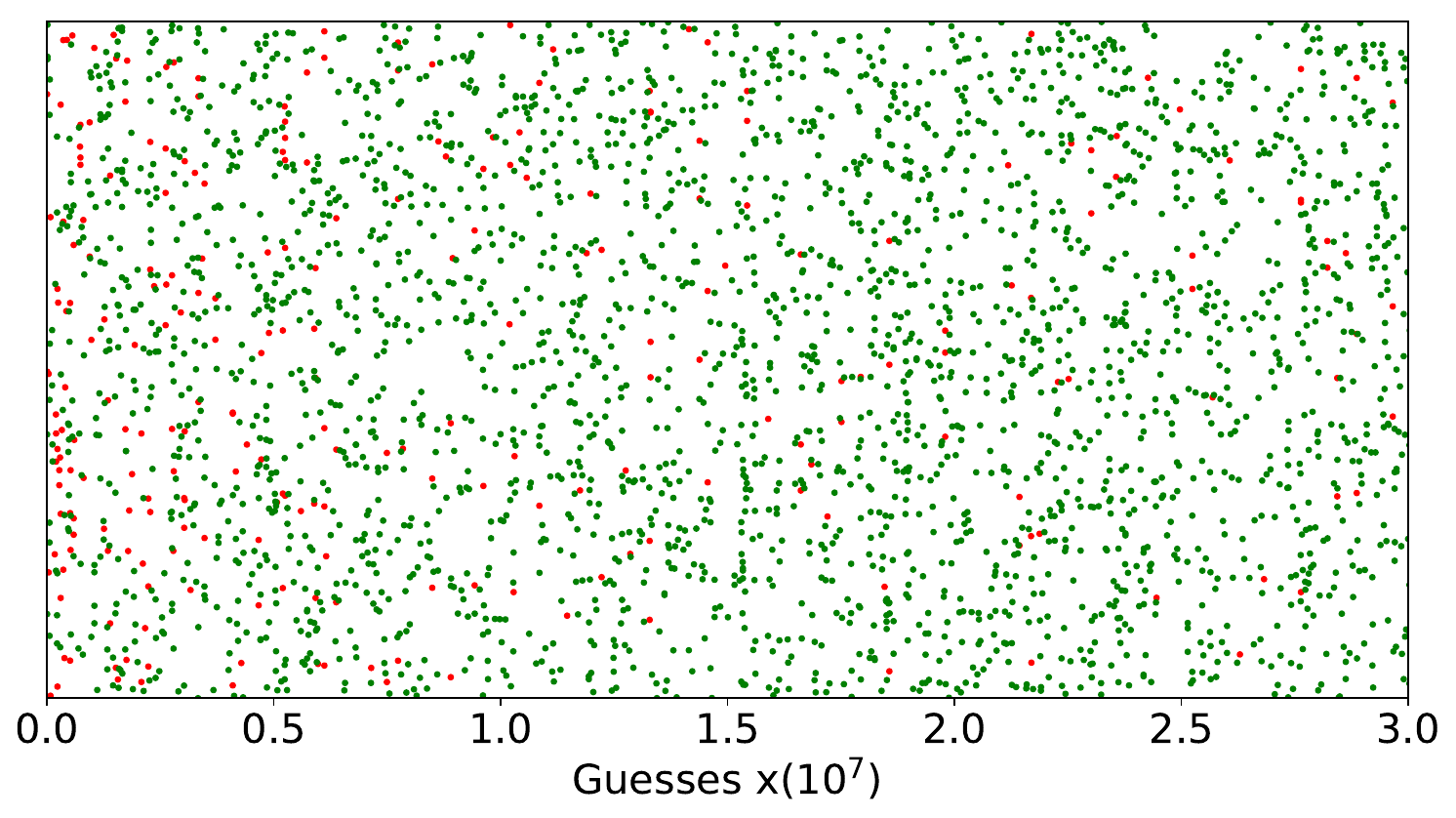}} 
    \subfigure[\texttt{5-order Markov (6-gram)}]{\includegraphics[width=0.3\textwidth]{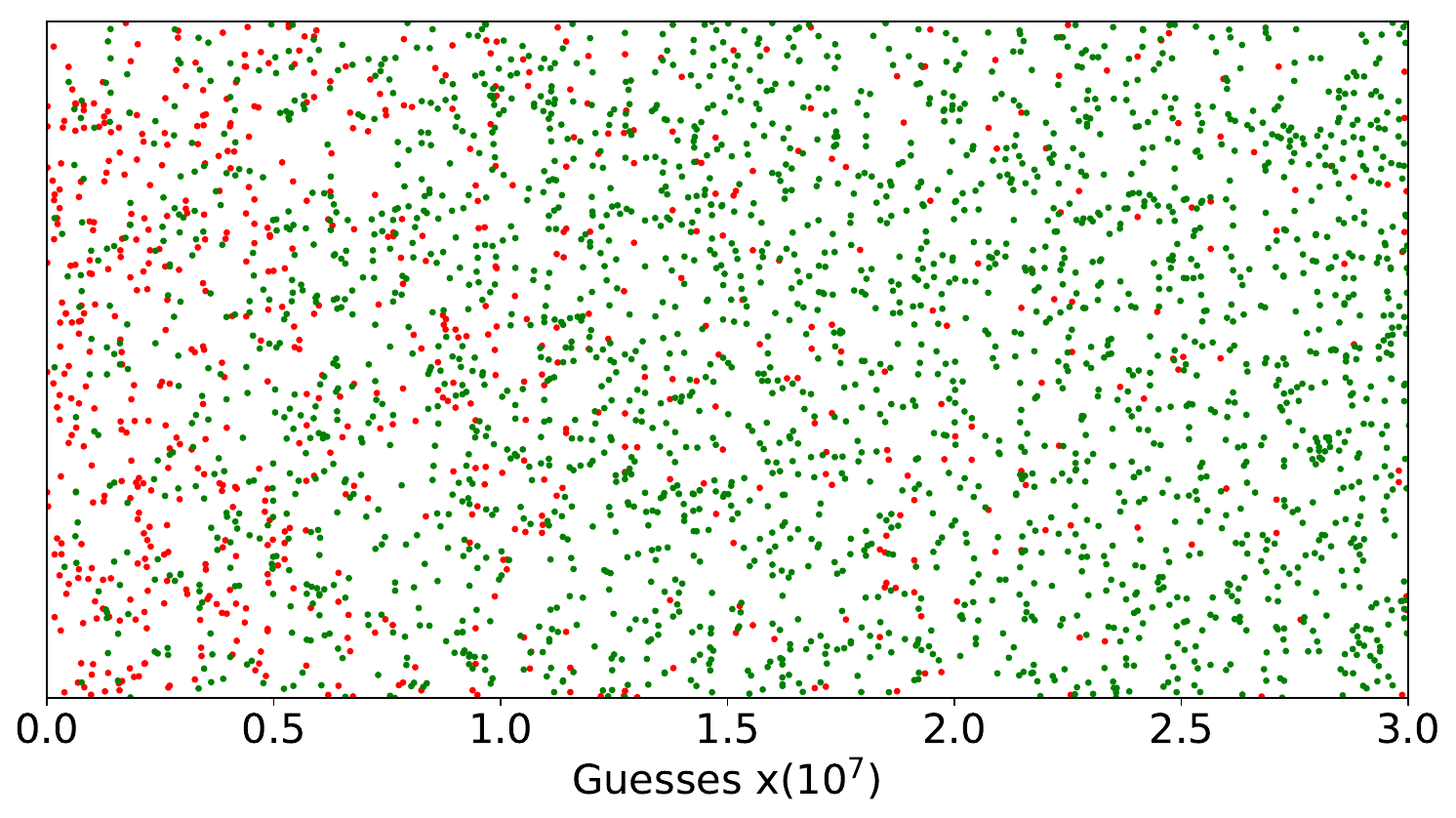}} 
    \subfigure[\texttt{7-order Markov (8-gram)}]{\includegraphics[width=0.3\textwidth]{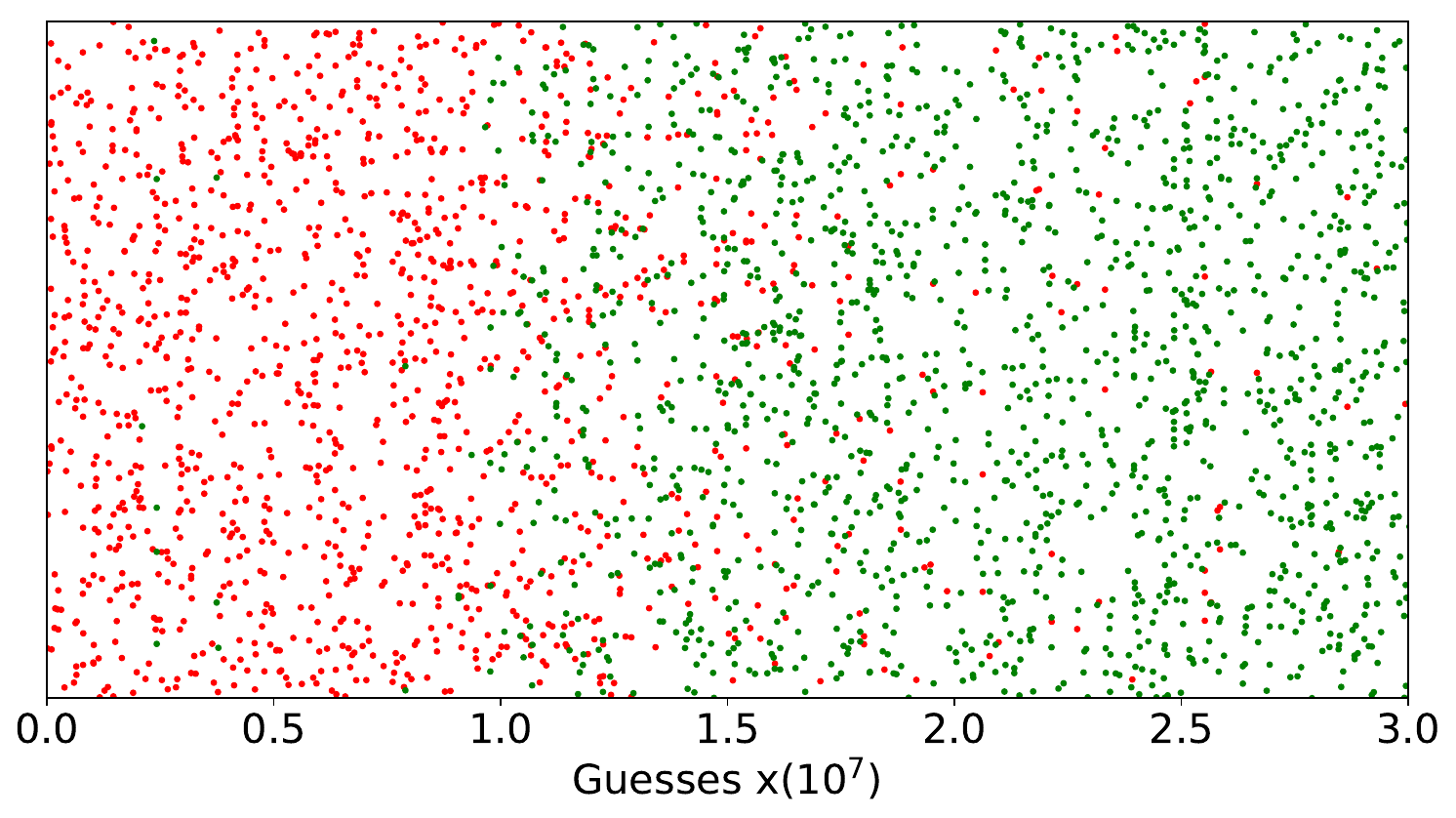}} 
    
    \caption{Over-learning manifestation across data-driven models, where areas with more red dots indicate severe over-learning phenomenon. We show more types of data-driven models on in Figure~\ref{fig:motivations:appendix} in Appendix~\ref{app:overlearning}.}
    \label{fig:motivations}
\end{figure*}

We train data-driven password models on \texttt{Rockyou} and evaluate them using samples from the training set (i.e., seen passwords) and from \texttt{XATO} (i.e., unseen passwords) with no overlap with the training data.
As shown in Figure~\ref{fig:motivations}, we use a scatter plot to show the distribution between trained passwords and non-trained passwords among those with fewer than $3\times 10^7$ guesses.
Each point in Figure~\ref{fig:motivations} is the evaluated password. The horizontal axis corresponds to the password's guess number as determined by the meter, with higher probabilities associated with smaller guesses. The vertical axis represents a random value. The color of a point indicates the member status, where the red one indicates a trained one, and the green one indicates a non-trained password. 
From Figure~\ref{fig:motivations}, the results reveal that the top candidates are mostly trained, suggesting that \textbf{a probability boundary can be established between member and non-member passwords.}

Furthermore, to quantify the extent of over-learning in data-driven meters, we introduce the metric $Fit_G$ to calculate their fitness to training data, which represents \textit{the percentage of member passwords among the top $G$ guesses}. A $Fit_G$ value nearing 1 indicates more pronounced over-learning and behavior akin to the List model.
We conduct training for each meter using the \texttt{Rockyou} dataset and generate guesses ranging from $G=10^4$ to $G=10^7$. The results of $Fit_G$ are presented in Table~\ref{table:fitness_accuracy}, revealing that the percentage of trained passwords decreases as the number of guesses $G$ increases. This is intuitive, as passwords with lower estimated probabilities are less likely to be included in the training data.
Our experimental findings also align with the fact that higher-order Markov models exhibit more severe over-learning issues, e.g., the 8-gram models' $Fit_{10^7}$ value is 0.96, indicating that the majority of generated passwords are member passwords.



\begin{table}[!htb]
\setlength{\abovecaptionskip}{0pt}
\setlength{\belowcaptionskip}{0pt}
\footnotesize
\centering
\caption{Fitness to trained passwords: $Fit_{G}$ refers the percentage of trained passwords found within the top $G$ guesses. The fitness to trained passwords decreases as the number of guesses increases.}
\label{table:fitness_accuracy}
\renewcommand\arraystretch{0.8}
\renewcommand\tabcolsep{11.5pt}
\begin{tabular}{c|cccc}
\toprule
\rowcolor{gray!45} Meters & $Fit_{10^4}$ & $Fit_{10^5}$ & $Fit_{10^6}$ & $Fit_{10^7}$\\ \midrule
V4.1\_PCFG  & 0.998      & 0.995 & 0.930    & 0.628                         \\
CKL\_PSM           & 0.998      & 0.970 & 0.651    & 0.228                         \\
\midrule
Backoff      & 0.999      & 0.998  & 0.816      & 0.431                     \\
4-gram      & 0.831      & 0.664  & 0.417      & 0.159                        \\
6-gram      & 0.984      & 0.945  & 0.798      & 0.429                     \\
8-gram      & 0.999      & 0.998  & 0.988      & \textbf{0.968}                      \\
AdaptivePSM       & 0.937      & 0.477  & 0.236    & 0.114                        \\  
\midrule
FLA PSM            & 0.995      & 0.986    & 0.765    & 0.275                      \\
\bottomrule
\end{tabular}
\end{table}

\subsection{Probability-threshold-selection Approaches} 
Based on the over-learning phenomenon, we first propose probability-threshold-selection approaches that aim to seek a probability threshold to distinguish between trained and non-trained passwords.

A simple way is to classify a password as trained if its estimated probability is greater than zero, that said, the threshold is zero. 
For the List model, the simple method is efficient enough, as the probability of identifying a trained password $P_{\mathit{in}}(x)=\frac{|D_{\mathit{train}}|}{|C_{\infty}(f_{target}) |} \sim 1$, where $|C_{\infty}(f_{target}) |$ refers to the generated candidate space of $f_{target}$, which equals to the count of all those passwords with a positive estimated probability. Then, $|C_{\infty}(f_{target})| = |D_{\mathit{train}}|$ in the case of the List model. In essence, candidate passwords that can be generated by the model typically yield a positive probability feedback. However, other data-driven password models have a better general-ability with $|C_{\infty}(f_{target})| \gg |D_{\mathit{train}}|$, signifying that $P_{in}(\mathit{x}) \sim 0$ for data-driven models.

To develop MIA approaches to target more models, we should carefully pick a probability threshold $\delta$, whose approach can be formally expressed as follows.

\begin{equation}\nonumber
\footnotesize
A(x) = \begin{cases}
\mathit{in}, & f_{target}(x) \ge \delta  \\
\mathit{out}, & f_{target}(x) < \delta
\end{cases}
\end{equation}

\noindent where $A(x)$ refers to the attack model that returns the member status of a password $x$ by the attack model, ``in'' and ``out'' refer to the member and non-member passwords, and $f_{target}(x)$ refers to the estimated probability of $x$ by the target meter.

\noindent\textbf{Finding the optimal threshold.}
To find the probability threshold, we resort to a shadow model denoted as $f_{\mathit{shadow}}$ that trained on the self-owned dataset using the same model architecture, to mimic the probability prediction behavior of the target model $f_{\mathit{target}}$. Using the shadow model, we can establish the mapping association between the percentage of trained passwords and a probability threshold, given their positive correlation.
It appears that the percentage of trained passwords declines 
with a decreasing threshold, as a password with low estimated probability is less likely to be included in the training sets. 
This association between the percentage of member passwords and the probability threshold can be extrapolated to the target model, assuming the shadow model accurately simulates the target model.
For brevity, we use the term \textbf{member ratio} to denote the proportion of member passwords among the queried passwords throughout the paper.

\begin{figure}[htb] 
\setlength{\abovecaptionskip}{0pt}
\setlength{\belowcaptionskip}{0pt}
\scalebox{1}{
\centering
\includegraphics[width=\linewidth]{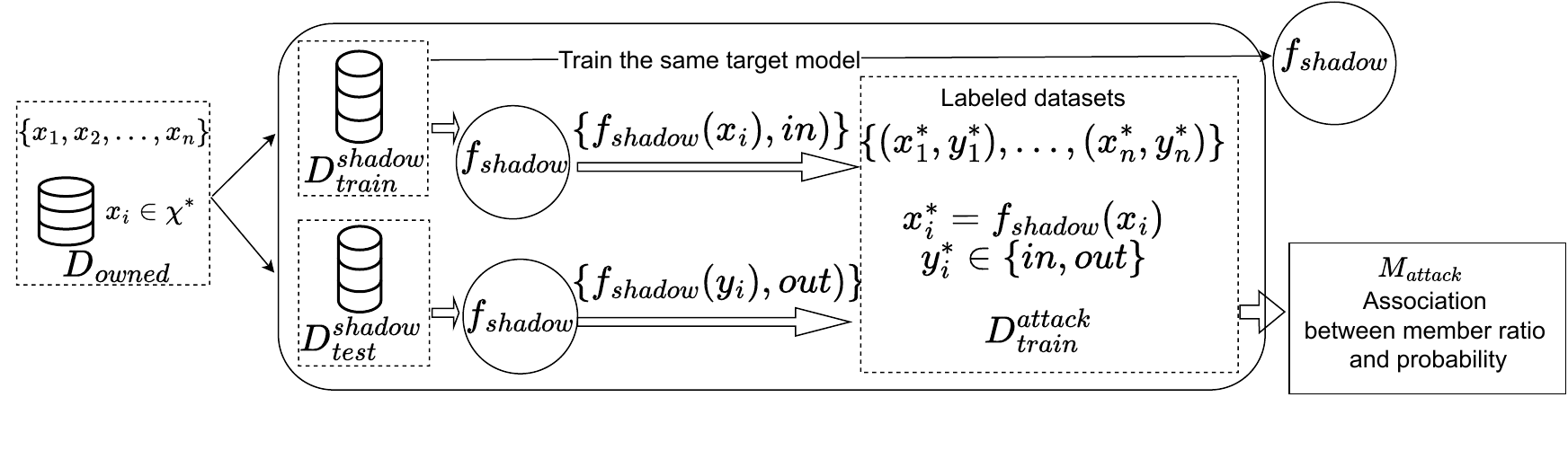}}
\caption{Overview of the probability-threshold-selection MIA method.}
\label{fig:methods}
\end{figure}

As a result, based on the association learned from the shadow model, our methodology is to settle down the probability threshold by a parameter of an expected member ratio we set. Specifically, as shown in Figure~\ref{fig:methods}, to train the shadow model and find the association, we randomly split the owned dataset $D_{\mathit{owned}}$ into two parts of $D_{train}^{shadow}$ and $D_{test}^{shadow}$ at 1:1 ratio, satisfying the $\forall x \in  D_{train}^{shadow}, x \notin D_{test}^{shadow}$. We utilize $D_{\mathit{train}}^{\mathit{shadow}}$ to train the shadow model $f_{\mathit{shadow}}$, while employing both $D_{train}^{shadow}$ and $D_{test}^{shadow}$ to query $f_{\mathit{shadow}}$ to obtain the labeled datasets. 
The labeled datasets comprise pairs of $x^*_i, y^*_i$, representing the estimated probability and the associate membership status respectively. We then sort the labeled results by descending order of probabilities $x^*_i$.
Based on the labeled results, we can determine the probability threshold $\delta$ according to an expected member ratio. Finally, we transfer the threshold to the target model, hoping that the target model can achieve the consistent member ratio. 
Then, we evaluate performance across multiple expected member ratios greater than 0.5, aiming for better performance compared to random guessing. 
This is based on the understanding that $f_{\mathit{shadow}}$ can replicate the behavior of $f_{\mathit{target}}$, and expect that such an accuracy can be achieved on $f_{\mathit{target}}$.


Finally, we highlight two key advantages of our approach. 
\begin{itemize}[fullwidth,itemindent=0em] 
    \item Flexibility: Attackers can adjust the expected member ratio to fit different attack scenarios. A higher member ratio can ensure the accuracy of MIAs, while a lower member ratio can add the number of identified trained passwords. 
    \item Efficiency: We also train a binary classifier based on features the labeled results (a set of $x^*_i, y^*_i$) from $f_{\mathit{shadow}}$. To fully investigate the potential of neural networks in this task,  we trained a binary classifier using three sets of features: the overall probability of a password, the internal probabilities of each token within the password, and a combination of these two features.
    However, our empirical results show that the binary classifier can only work better on neural-network-based meters, and the probability-threshold-selection methods generally yield better performance. 
    We leave more descriptions and results of the binary classifiers in Appendix~\ref{app:binary-classifer}. 
    
    We also note that Salem et al~\cite{DBLP:conf/ndss/Salem0HBF019ML-leaks} believed that it can weaken the assumption of the shadow model, and consider the top k\% records across the queried records of the target model as the members. They consider that $k=10$ can yield efficient membership inference attacks based on a large scale empirical studies. We also simulate the performance by regarding the top $k=10\%$ passwords as trained passwords when directly querying $D_{owned}$ to $f_{\mathit{target}}$. 
    
    To conserve space, we defer further comparisons between the most effective probability-threshold-selection method, binary classifier, and the direct application of Salem's method~\cite{DBLP:conf/ndss/Salem0HBF019ML-leaks} to Appendix~\ref{app:additional-results}. We find that these three approaches can also work against data-driven meters, and observe that the shadow model can significantly enhances MIAs. 
\end{itemize}

\subsection{Password Stealing Methods}     
To launch password stealing attacks, we need to employ generation models to generate additional passwords. GAN (Generative Adversarial Network) is an unsupervised generative model that learns the distribution based on adversarial learning. Hitaj~\cite{DBLP:conf/acns/HitajGAP19:GAN} et al. first applied GAN to password generation. Pasquini et al.~\cite{DBLP:conf/sp/PasquiniGABC21-representation-SP} proposed a dynamic GAN that improves the generation ability.

We leverage a GAN and a dynamic GAN to generate additional passwords, hoping to resemble the trained one, and then resort to a MIA'a attack model to distinguished trained password to construct the training database. Given a trained GAN model $G$, we formulate the process of generating more passwords as the following two steps. 
Firstly, we randomly sample points $z$ from the Gaussian distribution $\mathcal{N}(0, \sigma \mathbf{I})$. Second, the sampled points are fed into the generation model $G$ to yield passwords $x = G(z)$. This process allows us to generate additional passwords according to the condition $\{ x | x = G(z) \wedge p(z) \sim \mathcal{N}(0, \sigma \mathbf{I}) \}$.
Furthermore, the dynamic GAN~\cite{DBLP:conf/sp/PasquiniGABC21-representation-SP} could adjust the distribution of $G(z)$ based on the feedback to tailor the generated passwords more precisely. Specifically, when our membership inference attack model identifies a password as trained, we label the corresponding latent space $z_i$. Then, we replace the sampling distribution from random $\mathcal{N}(0, \sigma \mathbf{I})$ to $\mathcal{N}(z_i, \sigma \mathbf{I})$, facilitating the generation of additional passwords toward trained passwords.
We refer to the open-sourced model~\footnote{\url{https://github.com/pasquini-dario/PLR}} to train a GAN, in which the implementation of the generation adversarial network is Wasserstein GAN. On the basis of the static GAN, we use the most effective MIA approach to feedback the trained passwords to build the dynamic GAN. The generative model is composed of 5 layers of residual network with a dimension of 64 latent space, and an iteration round of 4,096.

\section{Evaluation~\label{sec:evaluation}}
\subsection{Experimental Configurations}
We illustrate the implementation details of these data-driven meters in Table~\ref{tab:surver-PSM}  as follows. 

\begin{itemize}[fullwidth,itemindent=0em]
    \item \textbf{AdaptivePSM:} We utilize a 4-gram Markov setting, as suggested in previous studies~\cite{kelley:guess, Ma:}.
    We follow the same parameters~\cite{DBLP:conf/ndss/CastellucciaDP12:adaptive-markov} including the noise value of $\gamma=5\cdot10^{-6}$ as the statistical frequency deviation. The noise mechanism refers to that every time a trained password is added to the Markov model, the statistical frequency is increased by one with a probability of $\gamma$. Therefore, a higher $\gamma$ value can provide more protection to the trained passwords.
    \item \textbf{FuzzyPSM:} FuzzyPSM uses the basic passwords as templates, and divides the password by the combination of a basic password and their variants of the basic password according to predefined rules. Following the prior work~\cite{DBLP:conf/dsn/WangHCW16:fuzzypsm}, we use phpBB~\footnote{\url{http://www. darkreading.com/risk/phpbb-password-analysis/d/d-id/1130335}} dictionary as the source for basic passwords. 
    \item \textbf{CKL\_PSM:} We use the publicly available password chunk dictionary~\footnote{\url{https://github.com/snow0011/CKL\_PSM}} to divide a password into several chunks and build the chunk-level PCFG for implementing CKL\_PSM.  
    \item \textbf{FLA:} We settle down a small-sized model with 200 LSTM cells, each containing a single hidden layer of 256 dimensions, as suggested in the open-sourced repository~\footnote{\url{https://github.com/cupslab/neural_network_cracking}} in client-side deployment. 
    \item \textbf{IPPSM:} We use the open-sourced auto-encoder implementation~\footnote{\url{https://github.com/pasquini-dario/InterpretablePPSM}} and maintain the same parameters including 10 CNN cells with hidden layers of 128 dimensions each.  
\end{itemize}

\noindent Further, to facilitate more intuitive comparison among meters of the same type, we also extend our evaluation to three additional models of Backoff~\cite{DellAmico:Monter_Carlo}, 4-gram Markov~\cite{DellAmico:Monter_Carlo} and V4.1\_PCFG~\cite{pcfg41}. This allows for the comparisons between AdaptivePSM and the default Markov-based models including Backoff and 4-gram, as well as, between FuzzyPSM, CKL\_PSM and V4.1\_PCFG, as they all fall under the PCFG-based category. We implement V4.1\_PCFG based on the open-sourced repository~\footnote{\url{https://github.com/lakiw/pcfg_cracker}} that divide a password with patterns such as keyboards to construct the template grammars.

\noindent\textbf{Experimental settings.} 
We train all the target data-driven models $f_{\mathit{target}}$ using the \texttt{Rockyou} dataset. 
We utilize \texttt{178} and \texttt{XATO} as the prior attack knowledge for $D_{owned}$ to train $f_{\mathit{shadow}}$, enabling to explore the impact of varying distribution of prior knowledge. The experimental settings simulate a realistic attack scenario, where an adversary acquires a newer dataset to infer information about a data-driven model trained on an older one. 
We further test the account security risks based on the newer dataset of \texttt{Collection\#1} and \texttt{Cit0day}.    
The leakage of information from an older dataset, such as \texttt{Rockyou-2009}, can still significantly influence newer password datasets, indicating the persistent impact for the attack despite evolving user behaviors. This risk becomes even more pronounced when simulating scenarios using more recent datasets. 

\subsection{Evaluation of Membership Inference Attacks~\label{sec:mia-evaluation}}

\noindent\textbf{Evaluation metrics.} 
MIAs entail a binary classification task, resulting in four possible outcomes: (1) True Positive (TP): the actual status is ``in'', and the estimated status is ``in''; (2) False Positive (FP): the actual status is ``out'', but the estimated status is ``in'' ; (3) True Negative (TN): the actual status is ``out'', and the estimated status is ``out''; (4) False Negative (FN): the actual status is ``in'', but the estimated status is ``out''.  
Based on these outcomes, we utilize precision, recall, and F1 score to evaluate the performance of MIAs.
\textbf{Precision} quantifies the proportion of correctly identified trained passwords among the passwords labeled as trained, calculated as $\frac{TP}{TP+FP}$.
\textbf{Recall} assesses the proportion of correctly identified trained passwords among all actual trained passwords, computed as $\frac{TP}{TP+FN}$. 
Here, trained passwords are the intersection between $D_{owned}$ and $D_{train}$, because attackers can only access the trained passwords they hold.
As a result, increasing the number of queried passwords typically enhance recall but may reduce precision. Solely depending on precision or recall for conclusive insights is inadequate. 
To strike a balance between the two, we introduce the F1 score to strike a balance between precision and recall. The F1 score, the harmonic mean of precision and recall, requires both to achieve higher values for F1 to increase.

\begin{figure}[htbp]
\footnotesize
\setlength{\abovecaptionskip}{0pt}
\setlength{\belowcaptionskip}{0pt}
  \centering
\includegraphics[width=0.46\textwidth]{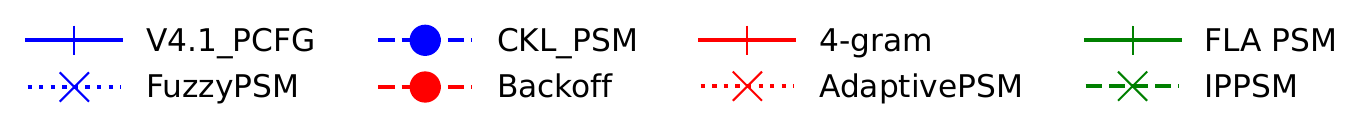} 
    \subfigure[\texttt{178}]{\includegraphics[width=0.23\textwidth]{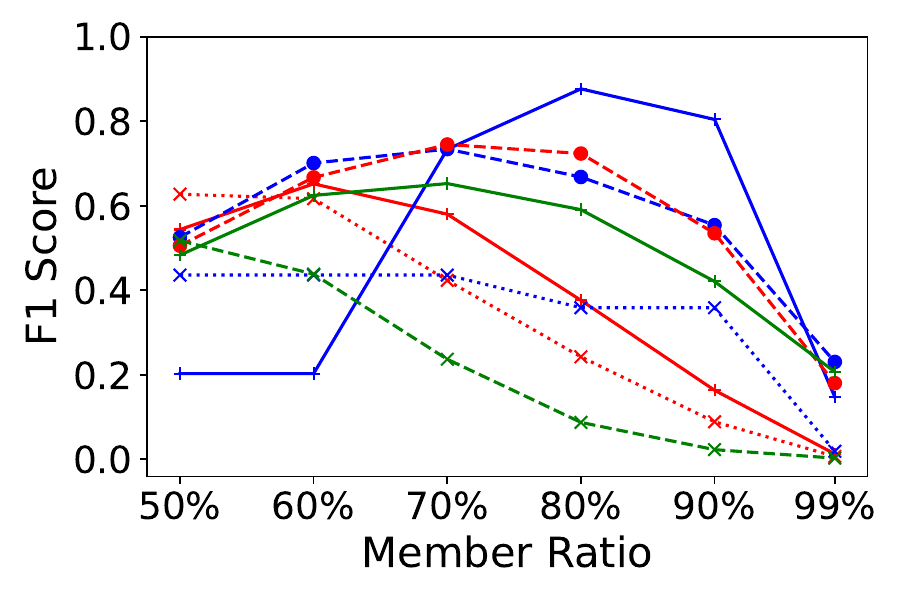}} 
    \subfigure[\texttt{XATO}]{\includegraphics[width=0.23\textwidth]{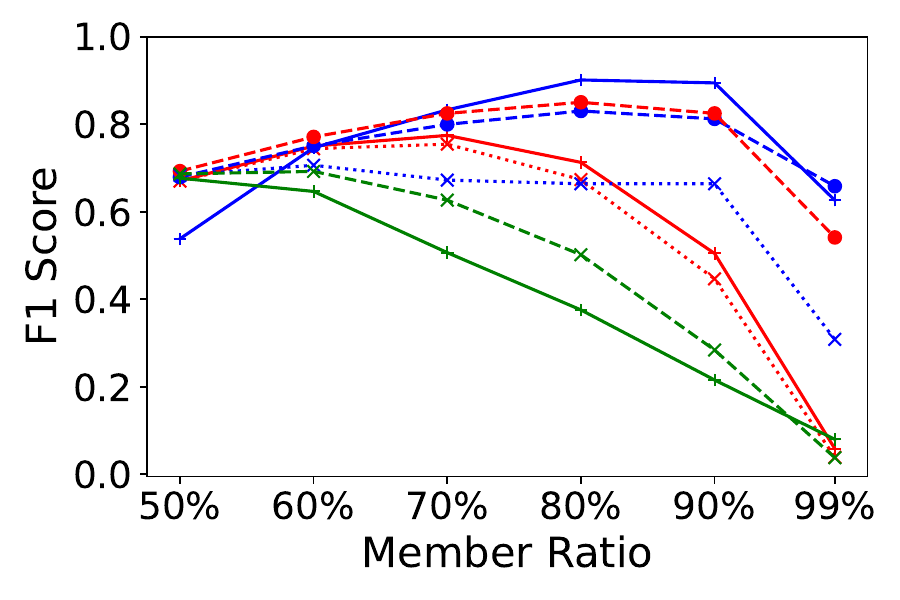}} 
    \caption{F1 scores of membership inference attacks under various expected member ratios.\label{fig:f1}}  
\end{figure} 

 
\noindent\textbf{Experimental results.} 
We primarily present F1 score under various expected member ratios exceeding 50\% in Figure~\ref{fig:f1}, and supplement the results of precision and recall in Figure~\ref{fig:percentile} in Appendix~\ref{app:additional-results}.
We conclude that membership inference attacks also effectively target several data-driven password meters by the high F1 scores achieved. The results increase first and then decrease, because a larger expected member ratio always denoting higher precision while lower recall.  
By this results, we suggest to use \textbf{the threshold associated with the 80\% expected member ratio} from $f_{\mathit{shadow}}$ to attack $f_{\mathit{target}}$ based on the accessible F1 score, coupled with relatively higher precision without a significant drop in recall.

We compare the effectiveness between our proposed three methods including the threshold associated with 80\% expected member ratio, binary classifier trained upon the combined features and the Salems' method of regarding the top 10\% of queried passwords as members~\cite{DBLP:conf/ndss/Salem0HBF019ML-leaks} in Figure~\ref{fig:ex_compare_mias} in Appendix~\ref{app:additional-results}. \textbf{We find that the three approaches can work effectively against data-driven meters in distinguishing member passwords, where the one based on the threshold associated with an 80\% member ratio performs better.} This method generally achieves higher F1 score than both the binary classifier and Salem's method, especially when knowledge of the attack is absent. Real-world attackers typically lack access to similar distributions to the actual training sets. 
Additionally, we also observe that the shadow model can enhance MIAs when comparing Salems' method in most cases.

\noindent\textbf{Experimental findings.} The neural-network-based meters are always robust against MIAs, Markov-based meters are the second, while PCFG-based meters are seriously vulnerable to MIAs. When compared about the Markov-based meters, AdaptivePSM is always more robust than the other two counter-parts of Backoff and 4-gram, demonstrating the effectiveness of the noise mechanisms in safeguarding the trained passwords.  
Moreover, the suffering of inference attacks is generally proportional to the over-learning of the model, as the PCFG-based models exhibit the most significant over-learning degree. 

Besides, MIAs are influenced by the distribution of prior attack knowledge. Specifically, we note that prior attack knowledge derived from an English dataset (\texttt{XATO}) tends to yield superior F1, precision, and recall scores compared to that derived from a Chinese dataset (\texttt{178}), indicating that attackers can typically achieve greater success when they possess attack knowledge that closely mirrors the distribution used to train the target model.

\noindent\textbf{Limitations.}   
We only use one set (i.e., \texttt{Rockyou}) to train the target models. 
However, we are confident that our approach is sufficient to draw convincing conclusions.
Firstly, we cover a broad spectrum of real-world scenarios including those with and without similar distributions. This ensures that our conclusions are broadly representative to a significant portion of real-world scenarios.
Furthermore, our focus is to assess the effectiveness of our proposed MIA methodology. We achieve this by employing a standardized and widely-used dataset (\texttt{Rockyou}) to train the target model and simulate various prior attack knowledge distribution.     

\subsection{Evaluation of Stealing Attacks}
In stealing attacks, we resort three experiments with different data sources for querying the target meter:
\begin{itemize}[fullwidth,itemindent=0em]
    \item Default: the owned dataset $D_\mathit{owned}$.   
    \item GAN and dynamic GAN: trained on $D_\mathit{owned}$ to generate additional passwords. The distinction lies in the dynamic GAN's ability to optimize the generation strategy based on feedback from MIAs (we settle down using the threshold associate to an 80\% expected member ratio). We configure both the GAN and dynamic GAN to generate $10^8$ passwords, matching the scale of a typical dataset $D_\mathit{owned}$ of around $10^8$ attack attempts. 
\end{itemize}

\noindent\textbf{Evaluation metrics.} 
We maintain a consistent high precision of 90\% and calculate the number of stolen passwords, ensuring a fair comparison.

\begin{table}[]
\setlength{\abovecaptionskip}{0pt}
\setlength{\belowcaptionskip}{0pt}
\footnotesize
 \renewcommand\arraystretch{1.2}
\renewcommand\tabcolsep{4.9pt} 
\centering
\caption{The number of stolen passwords in three stealing attacks. All these results are calculated with the same high precision of 90\%.} 
\label{tab:stealing_attack}
\begin{tabular}{c>{\columncolor{gray!45}}lll|>{\columncolor{gray!45}}lll|>{\columncolor{gray!45}}ll}
\toprule[\thickline]
\multicolumn{1}{c}{\multirow{2}{*}{Meters}} & \multicolumn{2}{c}{Default} &  & \multicolumn{2}{c}{Static GAN} &  & \multicolumn{2}{c}{Dynamic GAN} \\
 \multicolumn{1}{c}{}   & \texttt{178}          & \texttt{XATO}         &  & \texttt{178}            & \texttt{XATO}          &  & \texttt{178}            & \texttt{XATO}           \\ \cmidrule{1-3} \cmidrule{5-6} \cmidrule{8-9} 
V4.1\_PCFG  & 102,210 & 556,154        &  & +221\%  & +62\%         &  & +335\%          & 62\%      \\ 
FuzzyPSM & 40,920  &  268,678        &  & +183\%          & +60\%       &  &  +297\%        & +60\%   \\    
CKL\_PSM & 73,840 & 405,564      &  & +130\%         &  +9\%         &  & +210\%  & +9\% \\ 
\cmidrule{1-3} \cmidrule{5-6} \cmidrule{8-9}  
Backoff & 67,414 & 427,142 &  & +138\%          &  +53\%        &  & +318\%  & +53\%     \\
4-gram & 16,389 & 292,865 &  & +168\%          & +69\%       &  & +497\%       & +69\%      \\ 
AdaptivePSM & 8,213 & 248,955 &  & +158\%          & +47\%         &  & +639\%            & +47\%  \\ 
\cmidrule{1-3} \cmidrule{5-6} \cmidrule{8-9} 
FLA PSM & 44,862        & 104,998      &  & +133\% & +38\%        &  & +242\%          & +38\%       \\ 
IPPSM                                         & 1,042        & 76,647 &  & +15\%       & +18\%         &  &  +80\%       & +19\% \\
\bottomrule[\thickline]
\end{tabular}
\end{table}

\noindent\textbf{Experimental results.} 
We present the number of stolen passwords in Table~\ref{tab:stealing_attack}.
To more intuitively display the results, we illustrate the increased effectiveness of both the GAN and dynamic GAN over the default setting by displaying their percentage increase. For instance, if the GAN model extracts 120 trained passwords compared to 100 by the default method, this represents a 20\% increase.
From Table~\ref{tab:stealing_attack}, a significant improvement in the stolen numbers with GAN-based generation, with the dynamic GAN showing greater effectiveness in stealing trained passwords. 
Besides, the gain of dynamic GAN is closely tied to the distribution of prior attack knowledge. When attackers lack knowledge about the targeted distribution (i.e., experiments upon $D_\mathit{owned} = 178$), leveraging dynamic GAN to generate queried passwords can significantly enhances the stolen passwords.
Moreover, we observe that when a model is particularly vulnerable to MIAs, such as PCFG-based models, the stolen number tends to be consistently significant. Similarly, among Markov-based meters, AdaptivePSM demonstrates the best resilience against stealing attacks. 
Finally, in Appendix~\ref{app:stealing-upper-bound}, we conduct a stealing upper bound analysis where the number of stolen passwords typically ranges from $10^4$ to $10^5$ within a 90\% precision, as depicted in Figure~\ref{fig:ideal-stealing}, indicating the effectiveness of our empirical stealing methods.

\subsection{Evaluation Analysis}

\noindent\textbf{On the accuracy and security.} 
To evaluate meters' accuracy, we use the widely-used weighted spearman correlation coefficient~\cite{Golla:18ccspasswordmeter}. This coefficient compares the correlation between the ranks assigned by the meter and that of a testing dataset. The resulting value falls into [-1,1], with closer to 1 indicating better accuracy. We show the results in Table~\ref{tab:accuracy-security}, where we conclude that the accuracy and the security of data-driven meters are not a tradeoff. For example, FuzzyPSM achieves the best accuracy with a low level of suffering from MIAs, showcasing it is a better meter choice. In contrast, IPPSM has a low accuracy, but it suffer serious MIAs even than FuzzyPSM. 
Also, we find that the increasing of the noise $\gamma$ in AdaptivePSM is a tradeoff: a larger $\gamma$ reduces the leakage of trained passwords while decreasing accuracy.

\begin{tcolorbox}[colback=white!95!black,colframe=black!90!white,title= Accuracy and Security are not a Tradeoff, fonttitle=\bfseries, sharp corners=south] 
A low accuracy is not equal to a less leakage of trained passwords. Accuracy focuses on generalizing strength upon the universal password space $\pwddict$, while the security risk is caused by the over-learning of the trained passwords. 
\end{tcolorbox}

\begin{table}[]
\setlength{\abovecaptionskip}{0pt}
\setlength{\belowcaptionskip}{0pt}
\footnotesize
\renewcommand\tabcolsep{6.1pt} 
\centering
\caption{Comparison between accuracy of a meter and the security to resist membership inference attacks.} 
\label{tab:accuracy-security}
\begin{tabular}{clllll}
\toprule
 \multirow{2}{*}{Meters}                                   & \multicolumn{2}{c}{MIA's F1 score} &  & \multicolumn{2}{c}{Accuracy} \\ \cmidrule{2-3} \cmidrule{5-6} 
  & 178                 & XATO               &  & 178           & XATO         \\ \cmidrule{1-3} \cmidrule{5-6} 
V4.1\_PCFG                                                & 0.994               & 0.982              &  & 0.422         & 0.839        \\ 

\rowcolor{gray!45} FuzzyPSM                                                  & \textbf{0.820}               & \textbf{0.780}              &  & \textbf{0.502}         & \textbf{0.935}        \\ 
CKL\_PSM                                                  & 0.953               & 0.944              &  & 0.409         & 0.854        \\ 
\cmidrule{1-3} \cmidrule{5-6} 
Backoff                                                   & 0.954               & 0.908              &  & 0.471         & 0.871        \\
4-gram                                                    & 0.935               & 0.934              &  & 0.415         & 0.763        \\
\multicolumn{1}{l}{AdaptivePSM ($\gamma=5\times 10^{-4}$)} & 0.884               & 0.796              &  & 0.376         & 0.651        \\ 
\multicolumn{1}{l}{AdaptivePSM ($\gamma=5\times 10^{-5}$)} & 0.918               & 0.865              &  & 0.403         & 0.814        \\ 
\multicolumn{1}{l}{AdaptivePSM ($\gamma=5\times 10^{-6}$)} & 0.928               & 0.895              &  & 0.410         & 0.853        \\
\cmidrule{1-3} \cmidrule{5-6} 
FLA PSM                                                   & 0.898               & 0.905              &  & 0.350         & 0.863        \\ 
IPPSM                                                     & 0.829               & 0.869              &  & 0.218         & 0.019        \\ 
 \bottomrule
\end{tabular}
\end{table}

\begin{figure}[t]
\footnotesize
  \centering
    \subfigure[AdaptivePSM]{\includegraphics[width=0.13\textwidth]{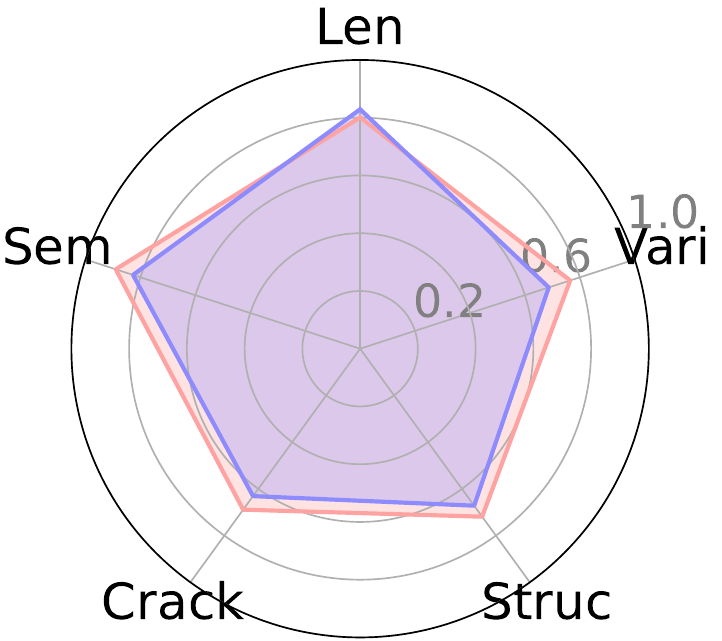}} 
    \subfigure[CKL\_PSM]{\includegraphics[width=0.13\textwidth]{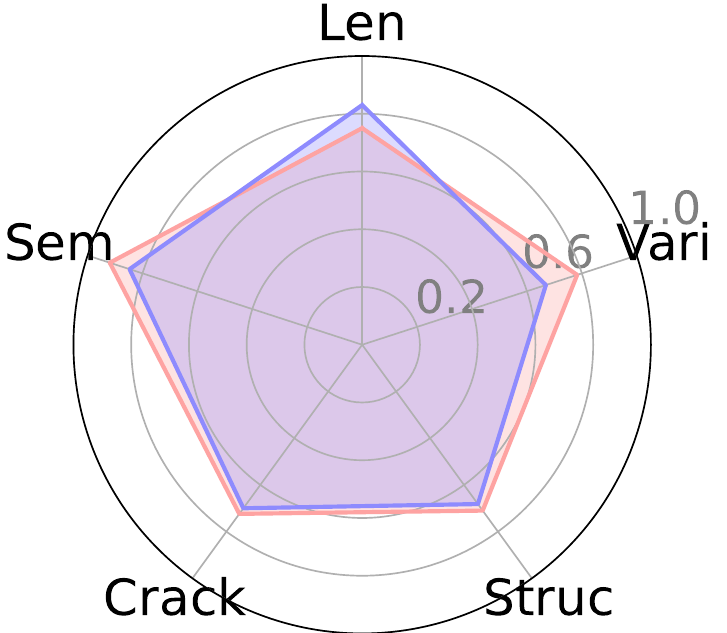}} 
    \subfigure[FLA]{\includegraphics[width=0.13\textwidth]{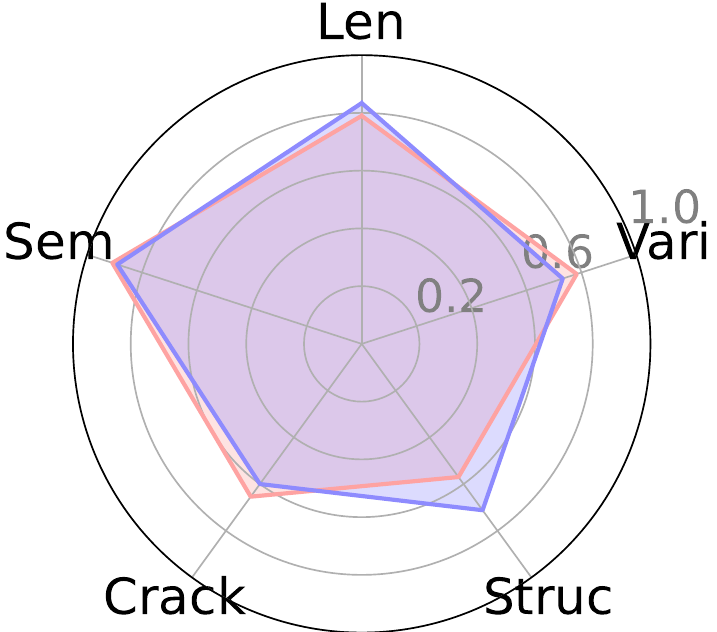}} 
    {\includegraphics[width=0.06\textwidth]{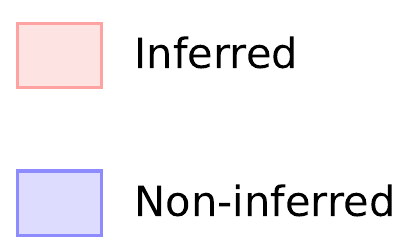}}
\caption{Radar chart of LLM-based evaluator for the characteristics of the inferred passwords.} 
\label{fig:radar}
\end{figure} 

\noindent\textbf{On the characteristics of stolen passwords.}  
An intriguing yet concerning aspect of the stolen passwords is their distinctiveness.
We use an LLM-as-a-evaluator~\cite{DBLP:conf/nips/ZhengC00WZL0LXZ23-LLM-judgement} to score the inferred passwords and the non-inferred passwords with the same size from the following perspectives.  Literature~\cite{DBLP:conf/nips/ZhengC00WZL0LXZ23-LLM-judgement} claimed that \emph{LLM-as-a-evaluator} is a salable and explainable way to approximate human preferences.

\begin{itemize}[fullwidth,itemindent=0em]
    \item \textbf{Semantic richness (Sem):} The LLM evaluates password semantics by recognizing natural-language words, keyboard patterns, word variants, assigning higher scores to passwords that are semantically rich.
    \item \textbf{Cracking resistance (Crack):} The LLM checks for the resistance to password cracking attacks, and gives higher scores for strong passwords.
    \item \textbf{Structure complexity (Struc):} The LLM checks for the structure complexity, ranging from the single character types, to mixed character types, and gives higher scores for complex structures. 
    \item \textbf{Variance (Vari):} The LLM checks for the variance and distinctiveness for the passwords, and gives higher scores for diverse passwords.
    \item \textbf{Length (Len):} The LLM compares the length and gives higher score for the longer lengths. 
\end{itemize}

We use the GPT-4o-mini for its effectiveness and affordability, and calculate the average scores based on AdaptivePSM, CKL\_PSM and FLA. We show the results in Redar figure~\ref{fig:radar}, where we can conclude that inferred passwords share similar characteristics with non-inferred ones, indicating that they resemble normal passwords rather than easily guessable ones, and are not trivially stolen. Particularly, the inferred passwords exhibit, e.g., more semantic richness, similar cracking resistance, less structure complexity, more variance/distinctiveness, and less length distribution.

Furthermore, 
We analyze the distribution of the stolen passwords (i.e., from dynamic GAN). We show the results in Table~\ref{tab:frequency-analysis}. 
Even low-frequency passwords, appearing in intervals of (0,10], still face risks of being stolen, with stolen rates as high as 41.1\% in V4.1\_PCFG. Based on these findings, we summarize the following conclusions.

\begin{tcolorbox}[colback=white!95!black,colframe=black!90!white,title=Low-frequency Passwords Suffer being Stolen, fonttitle=\bfseries, sharp corners=south] 
Passwords that are hard to crack, despite their diverse structures and semantics, can still be inferred during password-stealing attacks.
\end{tcolorbox}

\begin{table}[] 
\setlength{\abovecaptionskip}{0pt}
\setlength{\belowcaptionskip}{0pt}
\footnotesize
\renewcommand\tabcolsep{3.9pt} 
\centering
\caption{The percentage of trained passwords among the stolen ones at their frequency intervals.} 
\label{tab:frequency-analysis}
\begin{tabular}{clcl>{\columncolor{gray!45}}c>{\columncolor{gray!45}}ccc}
\toprule[\thickline]
\multirow{2}{*}{Meters}       &  & \multirow{2}{*}{Datasets} &  & \multicolumn{4}{c}{Percentage of member passwords} \\
                              &  &                           &  & (0,10{]}  & (10,$10^2${]}  & ($10^2$,$10^3${]}  & ($10^3$, $+\infty$ {]}  \\ 
                              \cmidrule{1-1} \cmidrule{3-3} \cmidrule{5-8} 
\multirow{2}{*}{V4.1\_PCFG}   &  & \texttt{178}                       &  &  29.5         &  99.7           & 99.8               &  99.9         \\
                              &  & \texttt{XATO}                      &  & 41.1 & 99.8             &    99.9           &   99.9         \\ 
                              \cmidrule{1-1} \cmidrule{3-3} \cmidrule{5-8}
\multirow{2}{*}{FuzzyPSM}     &  & \texttt{178}                       &  & 20.6          & 99.8            & 99.9              &  99.6          \\
                              &  & \texttt{XATO}                      &  & 71.6           & 99.9            & 99.9              & 99.9           \\ \cmidrule{1-1} \cmidrule{3-3} \cmidrule{5-8}
\multirow{2}{*}{CKL\_PSM}     &  & \texttt{178}                       &  & 30.8          &   53.1        & 29.5              &    25.5        \\
                              &  & \texttt{XATO}                      &  & 0.87          & 77.6       & 99.9              &  99.9          \\ 
\midrule[\thinline]
\multirow{2}{*}{Backoff}      &  & \texttt{178}                       &  & 9.53          & 99.9            & 99.9               &  99.6       \\
                              &  & \texttt{XATO}                      &  &    28.0       & 99.9            & 99.9              &  99.9          \\ \cmidrule{1-1} \cmidrule{3-3} \cmidrule{5-8} 
\multirow{2}{*}{4-gram}       &  & \texttt{178}                       &  & 3.69      & 32.6            & 70.9               &         93.6   \\
                              &  & \texttt{XATO}                      &  & 15.8           &     61.5  & 86.5              &  95.5          \\ \cmidrule{1-1} \cmidrule{3-3} \cmidrule{5-8}
  \multirow{2}{*}{AdaptivePSM} &  & \texttt{178}                       &  &  1.25 &     9.15   & 56.5      & 97.2           \\
                              &  & \texttt{XATO}                      &  & 7.59          & 42.2             &  79.7             &  99.4            \\ 
\midrule[\thinline]

\multirow{2}{*}{FLA}          &  & \texttt{178}                       &  & 1.61           &   80.0          & 97.8              &  99.6          \\
                              &  & \texttt{XATO}                      &  & 5.04          &           80.0  & 99.1              &  99.9          \\ \cmidrule{1-1} \cmidrule{3-3} \cmidrule{5-8} 
\multirow{2}{*}{IPPSM}        &  & \texttt{178}                       &  & 0.02          & 0.70            & 8.43              &  42.2         \\
                              &  & \texttt{XATO}                      &  & 2.50          &          11.6   & 38.1              &  70.8           \\ 
                              \bottomrule[\thickline]
\end{tabular}
\end{table}

\section{Account Security Risks~\label{sec:security-risks}} 
The undesirable disclosure of used passwords raises security concerns, limiting their practical utility. 
In this section, we propose a novel kind of threat model that leverages the used password leakage in Section~\ref{sec:targeted-guessing}.

\subsection{Novel Meter-aware Guessing Attacks~\label{sec:targeted-guessing}} 
\noindent\textbf{Threat model.} 
In a traditional targeted guessing scenario, an attacker attempts to crack a user's password based on their previously used passwords. If the attacker learns, possibly through social engineering, that the target account uses a password strength meter, they may focus on stealing the training data from the meter. By obtaining these used passwords, the attacker can significantly reduce the number of guesses needed and increase their chances of successfully compromising the user’s current password.

Targeted guessing~\cite{DBLP:conf/ccs/WangZWYH16,Das:passwordreuse, DBLP:conf/sp/PalD0R19:similarity} involves generating password candidates that closely resemble a user's previously used passwords, hitting the targeted passwords of other web services used by the same user.  
This principle lies in the tendency of users to reuse or slightly modify their historical passwords, thereby increasing the likelihood of successfully guessing passwords on other services used by the same individual.  
Formally, the targeted guessing can be described as 
$G(\widetilde{x}) \rightarrow x $ that leverages a previously used password $\widetilde{x}$ to hit the current used password $x$.

\begin{figure}[htb]
\setlength{\abovecaptionskip}{0pt}
\setlength{\belowcaptionskip}{0pt}
\centering
\setlength{\belowcaptionskip}{-4pt}
\scalebox{0.9}{
\includegraphics[width=\linewidth]{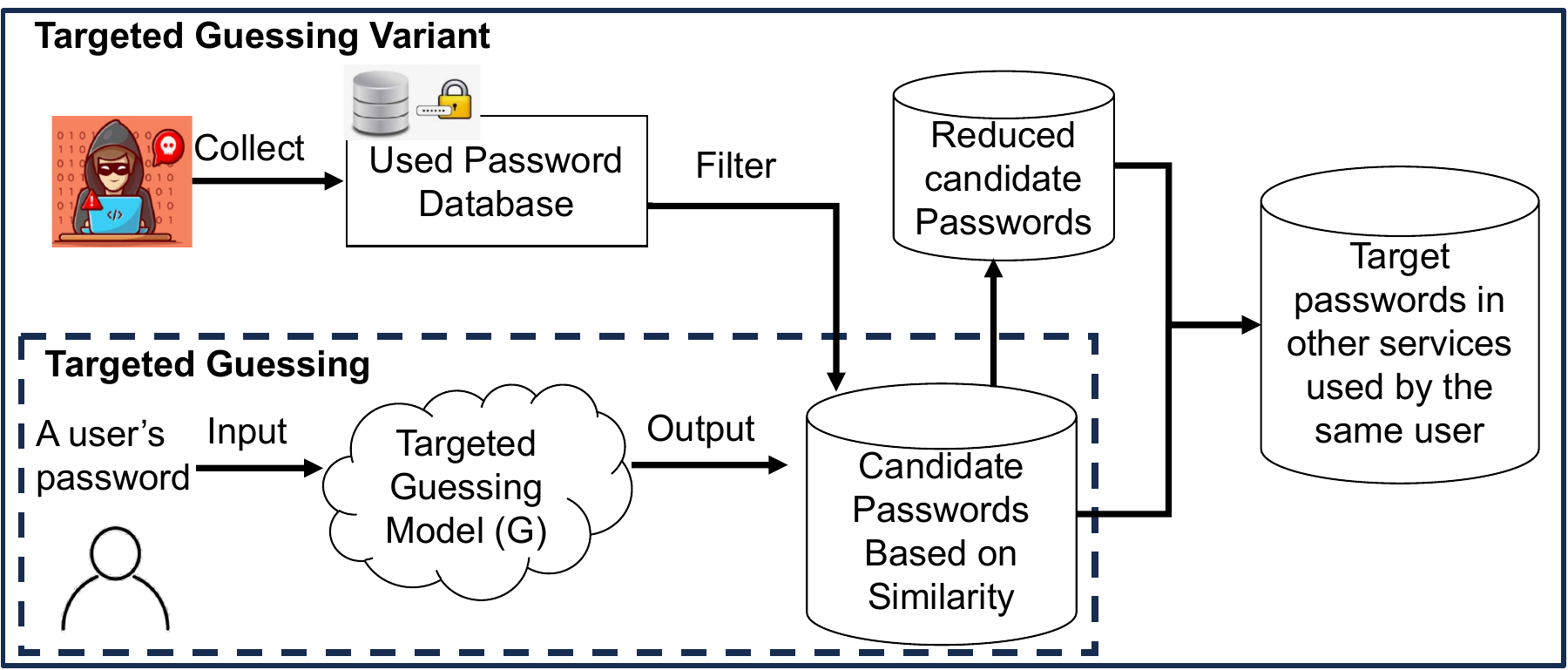}}
\caption{Meter-aware attack: sophisticated attackers filter the used passwords in their traditional targeted guessing scenarios.}   
\label{fig:targeted-guessing}
\end{figure}

A motivated attacker can further enhance targeted guessing by analyzing passwords used behind the meter, significantly increasing the cracking rate, posing serious risks to the account security of users relying on the meter.
As shown in Figure~\ref{fig:targeted-guessing}, adversaries will use this reconnaissance technique to aid attack operations, i.e., removing the used passwords of a meter, to improve the compromising abilities.
This filtering strategy can reduce the search space and accelerate targeted attacks, potentially leading to additional account compromises. This is because that users relying on the meter are less likely to choose the used passwords specific to the meter, as these are flagged as weak. 
For instance, websites often restrict the use of blocked or trained passwords due to their weak abilities, making such filtering strategies even more effective. 
Formally, the optimized targeted attack variant can be described as $G(\widetilde{x}, \widetilde{X}) \rightarrow x$, where $\widetilde{X}$ represents the collection of the used password dictionary by the website's meter.

Given our findings on the efficient trained password extraction attacks and the open disclosure of blocked passwords, adversaries can easily gather the used passwords behind the meter to optimize their attack strategies. Despite its apparent potential to improve the cracking rate based on our threat model, the magnitude of improvement achievable through these malicious activities remains unclear, which is pivotal.

\noindent\textbf{The used passwords specific to a meter.} 
In rule-based meters, we download three publicly accessible blocklist dictionaries from their open-sourced vendors: KeePSM~\footnote{\url{https://keepass.info/plugins.html}}, Zxcvbn~\footnote{\url{https://github.com/dropbox/zxcvbn}}, and CUPS PSM~\footnote{\url{https://github.com/cupslab/password\_meter}}. In data-driven meters, we regard the stolen passwords in Table~\ref{tab:stealing_attack} based on dynamic GAN experimented upon \texttt{XATO} as the trained passwords.  
We take the unique list of the used passwords in experiments.

\noindent\textbf{Experimental settings.}
We sample passwords of N users ($N_{\mathit{users}} = 10^5$) from \texttt{4iQ} and \texttt{Collection\#1} based on Emails. 
For an individual user, we then randomly pick two passwords~\cite{DBLP:conf/sp/PalD0R19:similarity}. We consider one of the passwords as leaked, and resort the well-known targeted model \emph{Pass2path}~\cite{DBLP:conf/sp/PalD0R19:similarity} to generate its variants as the candidate guesses.   
To avoid triggering account lockouts, we cap the number of guesses $g$ at $5$, $10$, and $100$.
If a match with the other password is found, it is considered as a successful hit, indicating a compromised account (labeled as hitted users). 
We then calculate the percentage of compromised users as $\frac{N_{\mathit{hitted \, users}}}{N_{\mathit{users}}}$.

\noindent\textbf{Experimental metrics.}
During the generation of guesses, an attacker skips any used passwords of the meter. In contrast, as a baseline for comparison, the attacker cannot exclude any guesses. Further, we also count how much accounts can be compromised earlier, and how many guesses can be averagely reduced in the whole cracking process with 100 guesses.   


\noindent\textbf{Experimental evaluation sets.}
As obtaining the leaked datasets from websites deploying these meters is not feasible, we make certain assumptions to construct evaluation sets for both rule-based and data-driven meters. For rule-based meters, we assume that datasets from such websites do not contain blocked passwords, as we can directly get the blocked status.  
For data-driven meters, we assume that datasets do not contain ``weak'' passwords labeled by the meter, as we can only get strength feedback. 
For example, to create evaluation sets for data-driven meters, we eliminate weak passwords with guesses smaller than $10^6$~\cite{DBLP:journals/cacm/FlorencioHO16:guess-strength} from the original evaluation sets such as \texttt{4iQ} and \texttt{Collection\#1}. 
This assumption is based on the understanding the meter has rejected the creation of such weak passwords for the website.
It's important to acknowledge that this assumption represents an idealized scenario for characterizing guessing scenarios.

\begin{table*}[]
\setlength{\abovecaptionskip}{0pt}
\setlength{\belowcaptionskip}{0pt}
\caption{Comparison between the attack performance upon rule-based meters with/without their blocklist dictionary.}
\label{tab:attackgains}
\renewcommand\tabcolsep{5.7pt}
\footnotesize
\begin{tabular}{clcllllcclccccc}
\toprule
 \multirow{3}{*}{Meters}          &  & \multirow{3}{*}{\begin{tabular}[c]{@{}l@{}}Whether to \\ leverage meter's\\ leaked dataset\end{tabular}} &  & \multicolumn{5}{c}{\texttt{4iQ}}                                                                                                                                                                                                             &  & \multicolumn{5}{c}{\texttt{Collection\#1}}                                                                                                                                                                                    \\ 
 \cmidrule{5-9} \cmidrule{11-15} &  & &  & \multicolumn{1}{c}{5} & \multicolumn{1}{c}{10} & 100     & \multicolumn{1}{l}{\begin{tabular}[c]{@{}l@{}}Earlier guessed \\ accounts\end{tabular}} & \multicolumn{1}{l}{\begin{tabular}[c]{@{}l@{}}Reduced\\  guesses\end{tabular}} &  & 5       & 10      & \multicolumn{1}{l}{100} & \multicolumn{1}{l}{\begin{tabular}[c]{@{}l@{}}Earlier guessed\\ accounts\end{tabular}} & \multicolumn{1}{l}{\begin{tabular}[c]{@{}l@{}}Reduced\\ guesses\end{tabular}} \\ \cmidrule{1-1} \cmidrule{3-3} \cmidrule{5-9} \cmidrule{11-15} 
 \rowcolor{gray!45} \cellcolor{white} \multirow{2}{*}{KeePSM}          &  &   yes                                                                                                      &  & 9.30\%                & 11.41\%                & 15.38\% & \multirow{2}{*}{12.33\%}                                                                & \multirow{2}{*}{2.31}                                                          &  & 10.30\% & 11.81\% & 15.88\%                  & \multirow{2}{*}{12.11\%}                                                               & \multirow{2}{*}{2.49} 
 \\  &  & no                                                                                                       &  & 9.14\%                & 10.86\%                & 15.34\%  &                                                                                         &                                                                                &  & 10.20\% & 11.63\% & 15.81\%                  &                                                                                        &                                                                               \\ \cmidrule{1-1} \cmidrule{3-3} \cmidrule{5-9} \cmidrule{11-15} 
\rowcolor{gray!45} \cellcolor{white} \multirow{2}{*}{Zxcvbn}          &  & yes                                                                                                      &  & 9.02\%                & 11.23\%                & 15.25\% & \multirow{2}{*}{16.12\%}                                                                & \multirow{2}{*}{3.29}                                                          &  & 10.30\% & 11.78\% & 15.91\%                 & \multirow{2}{*}{16.40\%}                                                               & \multirow{2}{*}{3.27}                                                         \\
                                 &  & no                                                                                                       &  & 8.80\%                & 10.61\%                & 15.20\% &                                                                                         &                                                                                &  & 10.19\% & 11.60\% & 15.79\%                &                                                                                        &                                                                               \\ \cmidrule{1-1} \cmidrule{3-3} \cmidrule{5-9} \cmidrule{11-15} 
\rowcolor{gray!45} \cellcolor{white} \multirow{2}{*}{CUPS PSM} &  & yes                                                                                                      &  & 8.93\%                & 11.10\%                & 15.13\% & \multirow{2}{*}{20.89\%}                                                                & \multirow{2}{*}{3.90}                                                          &  & 10.10\% & 11.52\% & 15.84\%                 & \multirow{2}{*}{20.23\%}                                                               & \multirow{2}{*}{5.07}                                                         \\
                                 &  & no                                                                                                       &  & 8.65\%                & 10.43\%                & 15.07\% &                                                                                         &                                                                                &  & 10.23\% & 11.73\% & 15.96\%                 &                                                                                        &                                                                               \\ \bottomrule
\end{tabular}
\end{table*}

\noindent\textbf{Experimental results.}   
We show the experimental results on websites employing three rule-based meters in Table~\ref{tab:attackgains}, and find that attackers can significantly cause additional account compromise in websites employing KeePSM, Zxcvbn and CUPS PSM. 
For instance, when provided with the Zxcvbn's blocklist dictionary, the percentage of compromised users increases by approximately 5.84\% 
with 10 guesses, highlighting the vulnerabilities for users registered on websites that have deployed Zxcvbn.
Besides, attackers can substantially expedite their attack success. 
For example, utilizing the Zxcvbn's blocklist, attackers can earlier compromise 16.12\% to 16.40\% of accounts, reducing the average attempted number by approximately 3 guesses.
We show experimental results of websites employing data-driven meters in Table~\ref{tab:ppsm-targeted-guessing}. The exposure of trained passwords from data-driven meters pose similar security threats as the blocked passwords: revealing the used passwords of a specific meter can cause additional compromises on accounts associated with websites that utilize this meter.  



\begin{table*}[!htbp] 
\setlength{\abovecaptionskip}{0pt}
\setlength{\belowcaptionskip}{0pt}
\caption{Comparison between the attack performance upon data-driven meters with/without their stolen trained passwords.} 
\label{tab:ppsm-targeted-guessing} 
\renewcommand\tabcolsep{5.7pt} 
\footnotesize
\begin{tabular}{clcllllcclccccc}
\toprule[\thickline]
\multirow{2}{*}{Meters}          &  & \multirow{2}{*}{\begin{tabular}[c]{@{}l@{}}Whether to \\ leverage meter's\\ leaked dataset\end{tabular}} &  & \multicolumn{5}{c}{\texttt{4iQ}}                                                                                                                                                                                                             &  & \multicolumn{5}{c}{\texttt{Collection\#1}}                                                                                                                                                                                    \\ \cmidrule{5-9} \cmidrule{11-15} 
                                 &  &                                                                                                          &  & \multicolumn{1}{c}{5} & \multicolumn{1}{c}{10} & 100     & \multicolumn{1}{l}{\begin{tabular}[c]{@{}l@{}}Earlier guessed \\ accounts\end{tabular}} & \multicolumn{1}{l}{\begin{tabular}[c]{@{}l@{}}Reduced\\  guesses\end{tabular}} &  & 5       & 10      & \multicolumn{1}{l}{100} & \multicolumn{1}{l}{\begin{tabular}[c]{@{}l@{}}Earlier guessed\\ accounts\end{tabular}} & \multicolumn{1}{l}{\begin{tabular}[c]{@{}l@{}}Reduced\\ guesses\end{tabular}} \\ \cmidrule{1-1} \cmidrule{3-3} \cmidrule{5-9} \cmidrule{11-15} 
\rowcolor{gray!45} \multirow{2}{*}{AdaptivePSM} \cellcolor{white}          &  & yes                                                                                                      &  & 9.07\%                & 11.25\%                & 15.37\% & \multirow{2}{*}{7.03\%}                                                                & \multirow{2}{*}{2.21}                                                          &  & 10.37\% & 11.91\% & 16.15\%                & \multirow{2}{*}{14.39\%}                                                               & \multirow{2}{*}{5.89}                                                         \\
                                 &  & no                                                                                                       &  & 9.06\%                & 10.85\%                & 15.43\% &                                                                                         &                                                                                &  &  10.36\% & 11.85\% & 16.10\% &                                                                                        &                                                                               \\ \cmidrule{1-1} \cmidrule{3-3} \cmidrule{5-9} \cmidrule{11-15} 
\rowcolor{gray!45} \multirow{2}{*}{CKL\_PSM} \cellcolor{white}          &  & yes                                                                                                      &  & 9.17\%                & 11.25\%                & 15.37\% & \multirow{2}{*}{16.58\%}                                                                & \multirow{2}{*}{4.45}                                                          &  & 7.12\% & 8.85\% & 14.15\%& \multirow{2}{*}{10.67\%}                                                               & \multirow{2}{*}{6.31}                                                         \\
                                 &  & no   &  & 8.92\%                & 10.85\%                & 15.48\% &                                                                                         &                                                                                &  & 6.93\% & 8.64\% & 14.06\%                 &                                                                                        &                                                                               \\ \cmidrule{1-1} \cmidrule{3-3} \cmidrule{5-9} \cmidrule{11-15} 
\rowcolor{gray!45} \multirow{2}{*}{FLA PSM} \cellcolor{white} &  & yes                                                                                                      &  &  9.02\%                & 11.27\%                & 15.43\% & \multirow{2}{*}{14.89\%}                                                                & \multirow{2}{*}{4.09}                                                          &  & 10.13\% & 11.61\% & 15.39\%                 & \multirow{2}{*}{8.46\%}                                                               & \multirow{2}{*}{5.07}                                                         \\
                                 &  & no  &  &  8.93\%                & 10.76\%                & 15.52\% & &                                                                                &  & 9.97\% & 11.59\% & 15.89\%                 &                                                                                        &                                                                               \\ \bottomrule[\thickline]
\end{tabular}
\end{table*}

\subsection{Security Consequences~\label{sec:sensitive-behaviors}}
We examine security consequences from the leakage of trained/blocked passwords.
We use the blocked passwords as an example to show some statistics. 
We consider the differences between a widely-used meter and a singly-deployed one, and regard the overlap among three blocklist dictionaries (KeePSM, Zxcvbn, and CUPS PSM) as indicative of a widely-used one.

First, leaked training passwords may contain personal identifiers, increasing the risk of linking them to specific user accounts. For instance, if the inferred password is ``Alice1997'', an attacker could reasonably deduce that it likely belongs to someone named Alice, potentially born in 1997. While such names are publicly known and may not, on their own, compromise an individual's privacy, their presence in public datasets introduces tangible risks. When an attacker is equipped with a large set of target users’ names, they could correlate these identifiers to enhance their success.
Specifically, we count the percentage of passwords that partially contain personal information including name, dates, and phone numbers. 
Further considering that users prefer to include self information into passwords~\cite{DBLP:conf/ccs/WangZWYH16,article-generating-and-remembering-passwords}, such dates and phone numbers are likely to be users' birthdays and related phone numbers.   
We conclude that the blocked passwords contain much name and date patterns, and less phone number patterns, and find a notable prevalence of names in a popular meter. 
We detail the algorithm for recognizing pattern passwords and the result in Table~\ref{tab:personal-percentage} in Appendix~\ref{app:proof}. 

\begin{figure}[htb]
\centering
\setlength{\belowcaptionskip}{-4pt}
\scalebox{1}{
\includegraphics[width=\linewidth]{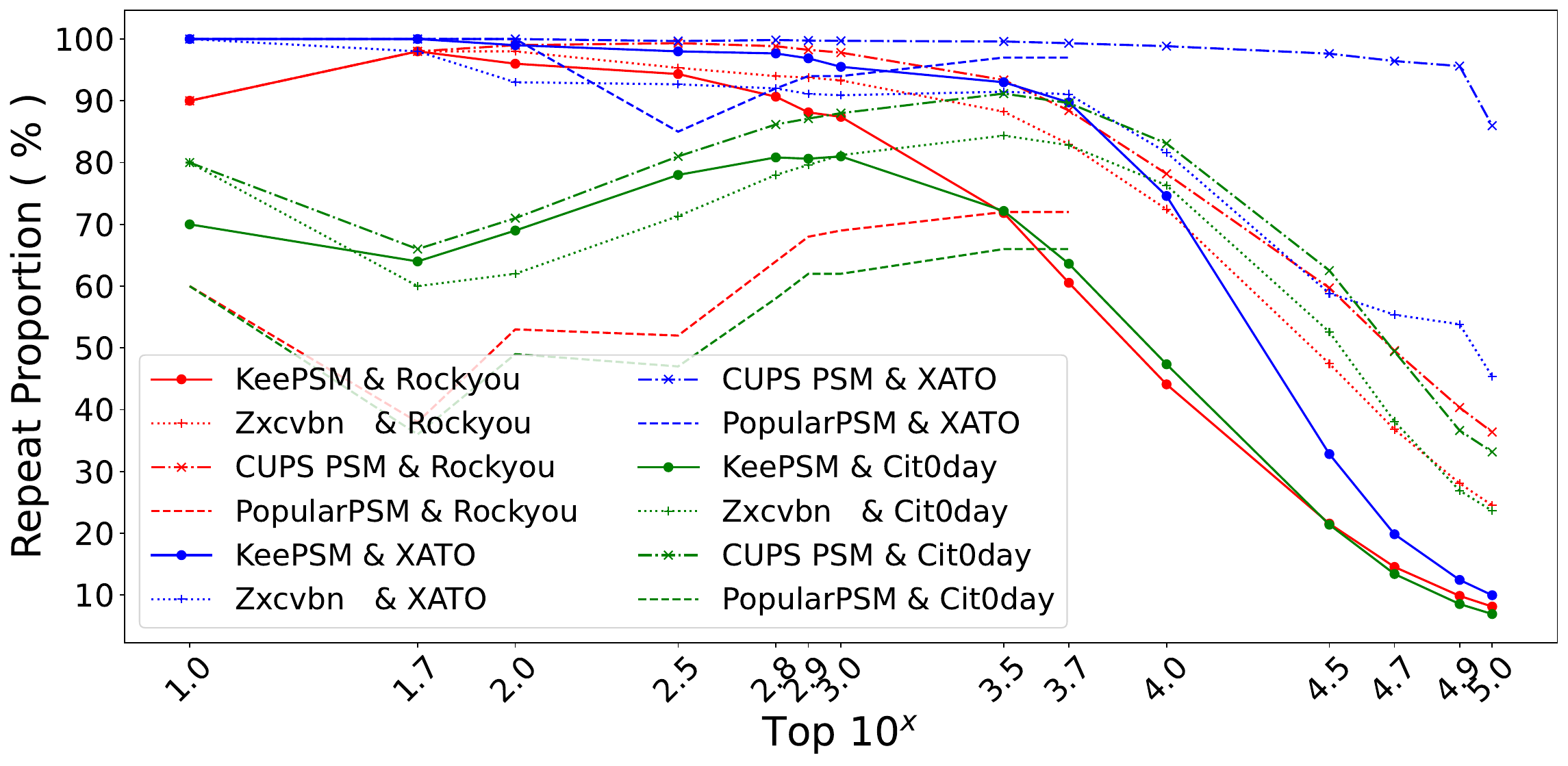}}
\caption{The repeat percentage between the top $10^x$ lists in blocked passwords and target datasets. The number of blocked passwords in a popular meter are limited, resulting in the line vanishing.} 
\label{fig:overlap}
\end{figure}

Second, the leaked passwords increase the credential compromise risks when users continue to reuse these stolen or blocked passwords on other accounts~\cite{Das:passwordreuse}.
Specifically, we compare the intersection proportion strictly between the top $10^5$ list of passwords in a blocklist dictionary and three general datasets. We show the results in Figure~\ref{fig:overlap}, where we can find that the blocklist dictionary of \texttt{KeePSM} and \texttt{Rockyou} (\texttt{KeePSM \& Rockyou}) have the 9 overlap passwords across the top 10 passwords.
We note a concerning trend where many users continue to reuse blocked passwords across different sites, with a high reuse rate for top passwords.
This could be because blocked passwords are often used on a single platform, leading users to reuse them across other services for convenience.
Fortunately, recent users in \texttt{Cit0day} shows a declining trend in the reuse of blocked passwords, suggesting that awareness and behaviors may be shifting positively over time.

\section{Discussion~\label{sec:discussion}}

\noindent\textbf{Defense counter-measures.}  
To protect the blocked passwords in a rule-based meter, we encourage to deploy a blocklist dictionary in the server-end and refrain it from the client-side. 
Relying on server-side defenses significantly strengthens protection by limiting an attacker's interaction with the system (e.g., fewer than 10 attempts), making it infeasible to infer meaningful information as MIAs typically require large-scale interactions (e.g., thousands of queries).
Several client-side protection mechanisms, such as simple data structures like Bloom filters, are insufficient, since attackers can download and analyze them offline to compromise the underlying database. 
The key challenge is that the server must additionally employ privacy-preserving protocols to ensure that users’ passwords are not exposed during the process. We leave communication cost optimization to future work but highlight initial protocol designs such as Google Password Checkup~\cite{DBLP:conf/uss/ThomasPYRKIBPPB19:GPC} and the FSB protocol~\cite{DBLP:conf/ccs/LiPAS0R19:FSB} as promising starting points.

Regarding the data-driven meters, we can leverage synthetic passwords~\cite{dong:DatasetCondensation}, e.g., generated by GAN, or differential privacy techniques to train the data-driven models, with the overhead optimization as future works.
Particularly, for the neural-network-based meters (i.e., FLA and IPPSM) that use the Softmax function to output probabilities, we recommend to increase the temperature ($T$) to reduce the probability difference. 
We formulate the probability $p_i$ for the predicted character $c_i$ as $p_i = \frac{exp(z_i/T)}{\sum_i{exp(z_i/T)}}$, where the $z_i$ refers to the output of the last layers of the neural network. $T$ and the probability difference are inversely proportional. We present the impact of $T$ to MIAs in Table~\ref{tab:T-impact}, where the $T=1$ represents to the original results. We conclude that increasing the temperature $T$ can generally mitigate the risks of MIAs, with the hyperparameter $T$ showing less impact for overall efficiency.

\noindent\textbf{Applicability of MIAs to Personalized PSMs.}   
Several personalized PSMs, such as PassBERT-PSM~\cite{xu-real-world-guessing} and PointerGuess-PSM~\cite{DBLP:conf/uss/Xiu024:pointerPSM}, have been proposed to estimate targeted guesses based on a leaked password—that is, how many attempts are needed to crack an input using a personalized model and its corresponding base password. These models typically learn reuse or transformation patterns trained on password pairs from the same user (e.g., linked via Email indicators). Our MIA method is suited for such personalized PSMs, leveraging shadow models to calibrate thresholds to distinguish member and non-member samples. These models exhibit pronounced differences in loss behavior between member and non-member samples, making it promising for the neural-network-based classier to identify optimal thresholds. 
The meter-aware threats from MIAs may be less applicable for personalized PSMs: personalized PSMs typically do not reject or limit reused passwords, making them less vulnerable to filtering-based attack threat. However, the security outcome for personalized PSMs becomes even severe, because their training data are often linked to email indicators.


\begin{table}[]
\setlength{\abovecaptionskip}{0pt} 
\setlength{\belowcaptionskip}{0pt}
\footnotesize
\renewcommand\tabcolsep{19.2pt} 
\centering
\caption{The impact of Temperature $T$ to the leakage of trained passwords upon neural-network-based meters.} 
\label{tab:T-impact}
\begin{tabular}{clll}
\toprule
 \multirow{2}{*}{Meter} & \multicolumn{1}{c}{\multirow{2}{*}{$T$}} & \multicolumn{2}{c}{Recall (\%)} \\    \cmidrule{3-4}  & \multicolumn{1}{c}{}                     & \texttt{178}          & \texttt{XATO}        \\ \midrule
\multirow{2}{*}{FLA PSM}   & $T=1$                                    &  44      & 12        \\
                & $T=1.5$                                  &  21         & 6        \\ \midrule
\multirow{2}{*}{IPPSM} & $T=1$                                    & 34         &  52        \\
 & $T=1.5$ & 30         & 54        \\ \bottomrule
\end{tabular}
\end{table}




\noindent\textbf{Takeaways.} 
(1) We demonstrate that trained password extraction attacks are feasible in prevalent public data-driven meters, which show the great potential of wider industrial applications. 
(2) We spotlight the security threat from real-world password strength meters, where the disclosure of the used passwords ubiquitous to a meter can cause additional account compromise from websites utilizing the meter. Ahead of potential attackers, we provide the defense counter-measure suggestions to mitigate these attacks, benefiting users and service providers.

\section{Conclusions~\label{sec:conclusion}}
This paper examines the effectiveness of public password strength meters in safeguarding used passwords that have been used in training or are on blocklists and concludes that they fall short in this regard. 
Through empirical and theoretical analyses, we find that data-driven meters leak between $10^4$ and $10^5$ trained passwords with high accuracy. 
Furthermore, we illustrate a security threat where the exposure of trained or blocked passwords of a meter can lead to additional compromises on accounts associated with the meter, demonstrating the need for the development of robust and privacy-preserving meters.
Our findings reveal that password strength meters are potentially vulnerable gatekeepers in website security.  
To mitigate these vulnerabilities, we propose the defense suggestions to enhance meter robustness with privacy-preserving mechanisms.






\bibliographystyle{plain}
\bibliography{reference}

\appendix

\section{Additional Analysis, Approaches and Results of Password Inference and Stealing Attacks}

\subsection{More Results of Over-learning Phenomenon}~\label{app:overlearning}
\begin{figure*}[!htb]
    \centering
    \subfigcapskip=-6pt
        \subfigure[\texttt{Backoff}]{\includegraphics[width=0.3\textwidth]{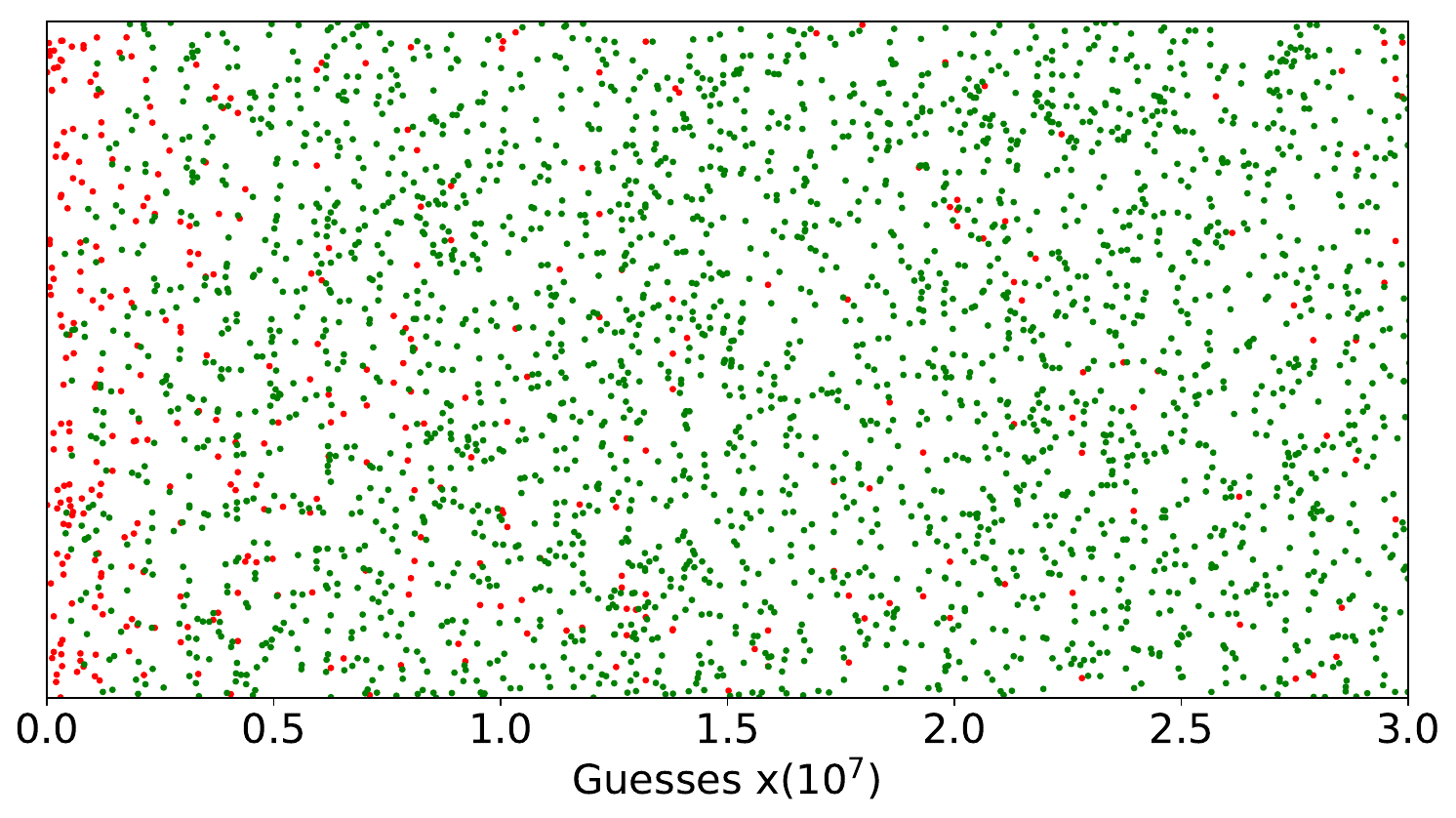}} 
    \subfigure[\texttt{FLA}]{\includegraphics[width=0.3\textwidth]{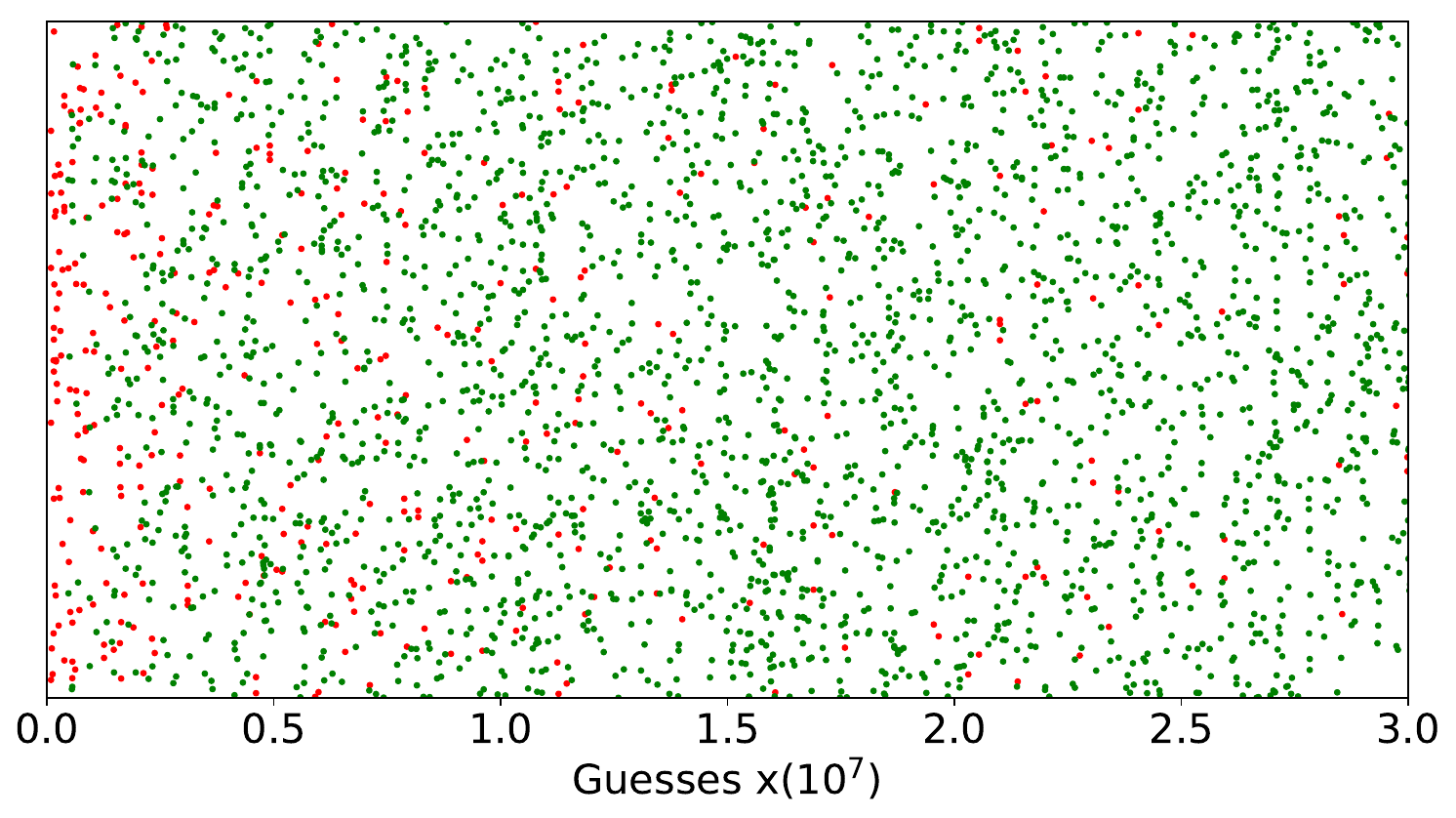}}
    \subfigure[\texttt{V4.1\_PCFG}]{\includegraphics[width=0.3\textwidth]{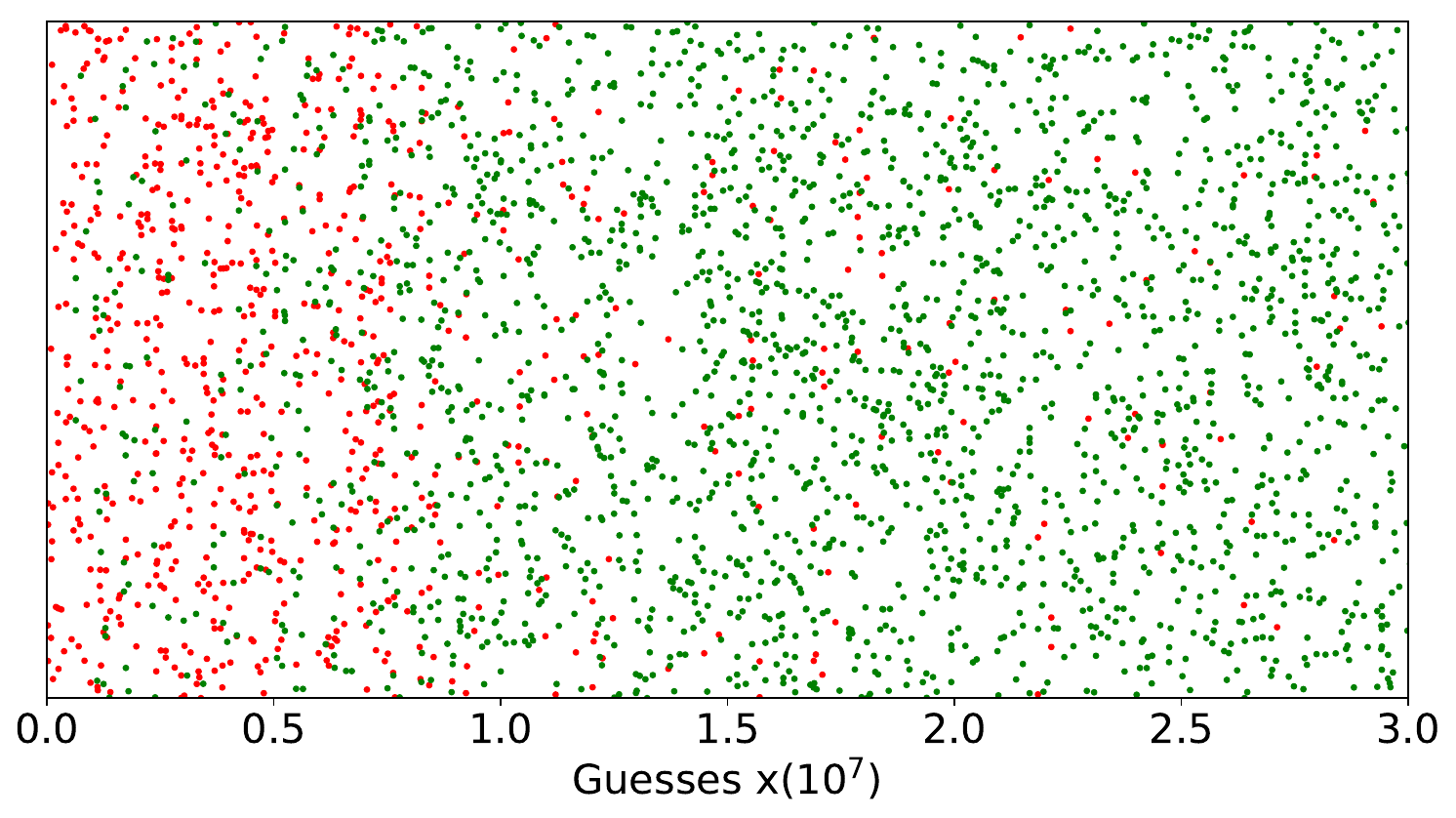}}   
    \caption{Over-learning manifestation across more types of data-driven models. Areas with more red dots indicate severe over-learning phenomenon.}
    \label{fig:motivations:appendix}
\end{figure*}

We show the over-learning phenomenon upon more data-driven meters in Figure~\ref{fig:motivations:appendix}, which exhibits consistent over-learning phenomenon in n-grams (shown in Figure~\ref{fig:motivations}).

\subsection{Binary-classifier MIA Approach}~\label{app:binary-classifer}
Based on the results of the labeled dataset of $$D^{*} = \{[f_{shadow}(x_1), \mathit{in}], [f_{shadow}(x_2)] , \mathit{out}),\dots  \}$$ from shadow model, we can build a binary-classifier to learn the distinction between member passwords and non-member passwords and distinguish them better. The ``in'' and ``out'' denote the member and non-member status. 
We build the binary classifier with the following features (i.e., supervised manner with a feature labeled with the membership status).  
\begin{itemize}[fullwidth,itemindent=0em]
    \item $P(x|M)$: the feature is the whole probability estimated from the password.
    \item $I(x|M)$: the feature is the collection of internal probabilities of every token in a password, for example, the character's probability in Markov-based and neural-network-based meters, and the template's probability in PCFG-based meters. 
    \item $J(x|M)$: the feature is the whole probability and the internal probabilities. 
\end{itemize}
We can explore the effect of these different features to the MIA performance. 
Given that the recurrent neural networks can better handle input data of variable lengths, we consider recurrent neural networks suitable in this scenario. 
Specifically, we use the LSTM as the recurrent neural network with a hidden layer of 256 layers. We resort 3 LSTMs and use the cross-entropy as the objective function. We use PyTorch framework and train 20 iterations.

We show the results of the binary classifiers based on the respective three features in Table~\ref{table:experiment_binary_classifier}. We find that the jointed features ($J(x|M)$) can generally yield better attack gains in F1 score in most of the meters. 
Besides, the binary classifier can generally yield a little attack gains than the threshold-selecting approaches for neural-network-based meters, while threshold-selecting approaches can work efficiently for Markov-based and PCFG-based meters.

\begin{table*}[!htb]
\setlength{\abovecaptionskip}{0pt} 
\setlength{\belowcaptionskip}{0pt}
\footnotesize
\centering
\caption{The experimental results based on three binary classifiers with three features.}
\label{table:experiment_binary_classifier}
\renewcommand\arraystretch{1}
\resizebox{0.75\textwidth}{!}
{
\begin{tabular}{c|c|ccc|ccc}
\toprule
\multirow{2}{*}{Meters}       & \multirow{2}{*}{Features} & \multicolumn{3}{c|}{\texttt{178}} & \multicolumn{3}{c}{\texttt{XATO}} \\ \cmidrule{3-8} 
                               &                       & F1    & Precision    & Recall    & F1    & Precision    & Recall    \\ \midrule
\multirow{3}{*}{V4.1\_PCFG} & $P(w|M)$              & 0.914  & 0.885  & 0.946  & 0.874  & 0.980  & 0.789  \\
                               & $I(w|M)$              & 0.637  & 0.746  & 0.556  & 0.838  & 0.822  & 0.854  \\
                               & $J(w|M)$              & 0.822  & 0.769  & 0.882  & 0.878  & 0.968  & 0.803  \\ \midrule

\multirow{3}{*}{FuzzyPSM}      & $P(w|M)$              & 0.254  & 0.983  & 0.146  & 0.693  & 0.800  & 0.612  \\
                               & $I(w|M)$              & 0.024  & 0.901  & 0.012  & 0.006  & 0.623  & 0.003  \\
                               & $J(w|M)$              & 0.253  & 0.983  & 0.145  & 0.710  & 0.804  & 0.636  \\ \midrule
\multirow{3}{*}{CKL\_PSM}      & $P(w|M)$              & 0.486  & 0.974  & 0.324  & 0.730  & 0.989  & 0.579  \\
                               & $I(w|M)$              & 0.035  & 0.788  & 0.018  & 0.630  & 0.921  & 0.479  \\
                               & $J(w|M)$              & 0.479  & 0.976  & 0.317  & 0.727  & 0.990  & 0.575  \\ \midrule
\multirow{3}{*}{Backoff}       & $P(w|M)$              & 0.531 & 0.977  & 0.365  & 0.846  & 0.901  & 0.797  \\
                               & $I(w|M)$              & 0.653  & 0.795  & 0.554  & 0.833  & 0.901  & 0.774  \\
                               & $J(w|M)$              & 0.671  & 0.870  & 0.546  & 0.810  & 0.883  & 0.748  \\ \midrule
\multirow{3}{*}{4-gram}         & $P(w|M)$              & 0.452 & 0.886  & 0.303  & 0.728  & 0.847  & 0.639  \\
                               & $I(w|M)$              & 0.484  & 0.654  & 0.384  & 0.743  & 0.830  & 0.673  \\
                               & $J(w|M)$              & 0.456 & 0.745  & 0.329  & 0.736  & 0.846  & 0.652  \\ \midrule
\multirow{3}{*}{Adaptive PSM}  & $P(w|M)$              & 0.272  & 0.938  & 0.159  & 0.698  & 0.851  & 0.592  \\
                               & $I(w|M)$              & 0.343  & 0.675  & 0.230  & 0.728  & 0.826  & 0.651  \\
                               & $J(w|M)$              & 0.377  & 0.740  & 0.253  & 0.737  & 0.853  & 0.648  \\ \midrule                         
\multirow{3}{*}{FLA PSM}       & $P(w|M)$              & 0.525  & 0.809  & 0.389  & 0.161  & 0.866  & 0.089  \\
                               & $I(w|M)$              & 0.547  & 0.759  & 0.427  & 0.204  & 0.836  & 0.116  \\
                               & $J(w|M)$              & 0.523  & 0.780  & 0.393  & 0.234  & 0.873  & 0.135  \\ \midrule
\multirow{3}{*}{IPPSM}         & $P(w|M)$              & 0.254  & 0.754  & 0.153  & 0.591  & 0.781  & 0.475  \\
                               & $I(w|M)$              & 0.459  & 0.762  & 0.328  & 0.631  & 0.785  & 0.527  \\
                               & $J(w|M)$              & 0.504  & 0.754  & 0.379  & 0.627  & 0.836  & 0.501  \\ \bottomrule
\end{tabular}
}
\end{table*}

\subsection{Additional Comparison Results of Membership Inference Attacks}~\label{app:additional-results}

\begin{figure*}[h]
\setlength{\abovecaptionskip}{0pt} 
\setlength{\belowcaptionskip}{0pt}
\footnotesize
\setlength{\abovecaptionskip}{0pt}
\setlength{\belowcaptionskip}{0pt}
\centering
\includegraphics[width=\linewidth]{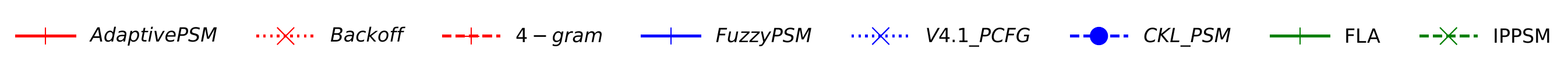}%
\\
\subfigure[Precision (\texttt{178})]{\includegraphics[width=0.23\textwidth]{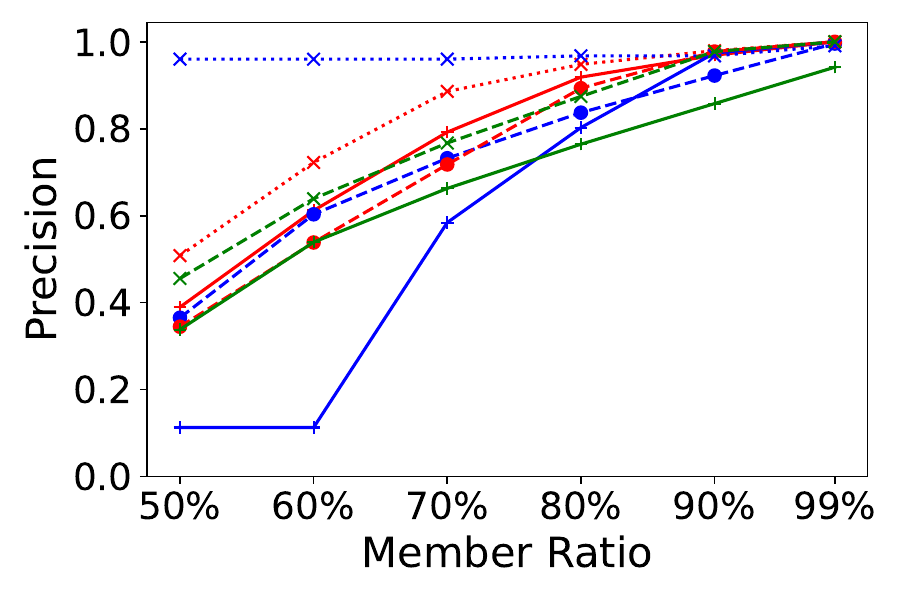}} 
\subfigure[Recall (\texttt{178})]{\includegraphics[width=0.23\textwidth]{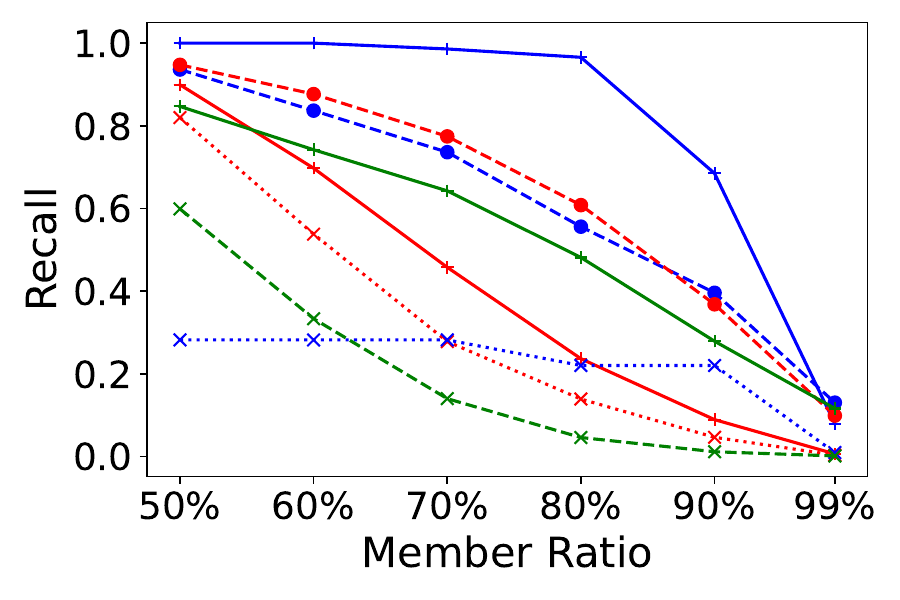}} 
\subfigure[Precision (\texttt{XATO})]{\includegraphics[width=0.23\textwidth]{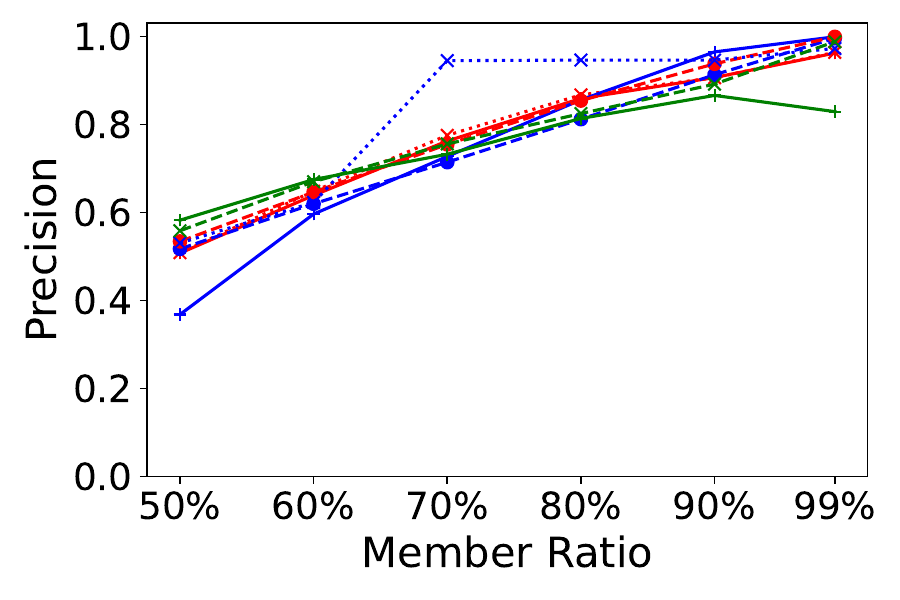}} 
\subfigure[Recall (\texttt{XATO})]{\includegraphics[width=0.23\textwidth]{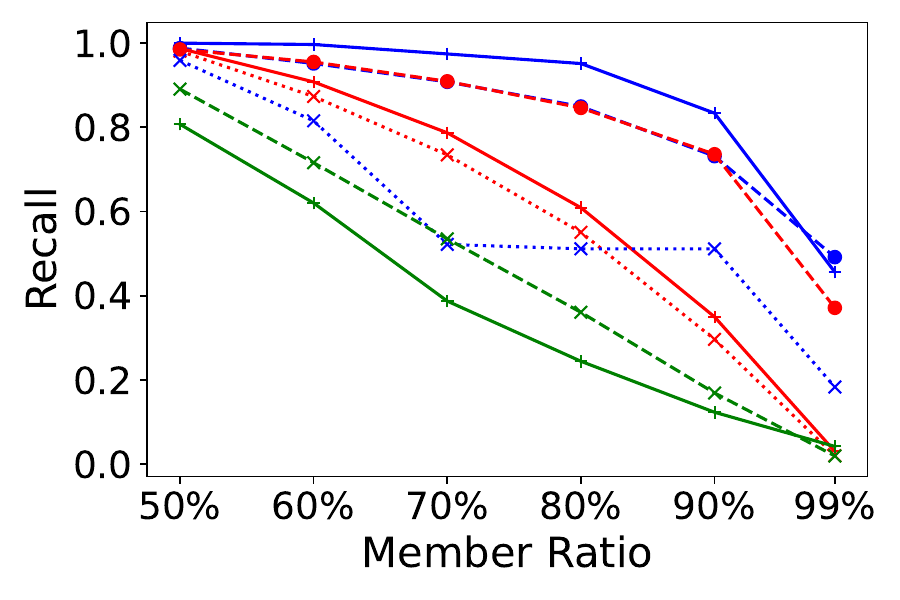}} 
\caption{Precision and recall of MIAs.} 
\label{fig:percentile}
\vspace{-3mm}
\end{figure*}

We show precision and recall associated with each expected precision in Figure~\ref{fig:percentile}. 
Similar with previous observations, we find that AdaptivePSM consistently exhibits lower scores. 
We also note that the recall scores are extremely small with a high threshold, especially when trained upon \texttt{178} in a shadow model. This is due to the small overlap (4,094) between \texttt{178} and \texttt{Rockyou} on top $10^5$ passwords, compared with that of 39,606 with \texttt{XATO}.

\begin{figure*}[!htbp]
\setlength{\abovecaptionskip}{0pt} 
\setlength{\belowcaptionskip}{0pt}
    \centering
\includegraphics[width=0.65\linewidth]{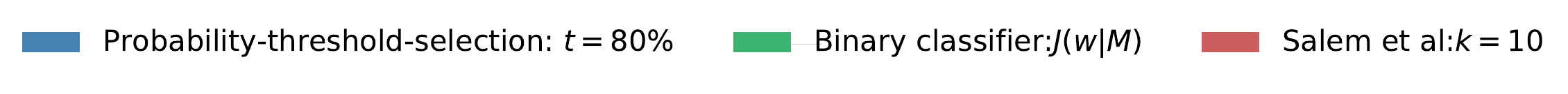}
    \subfigure[F1 (\texttt{178})]{
        \includegraphics[width=0.45\linewidth]{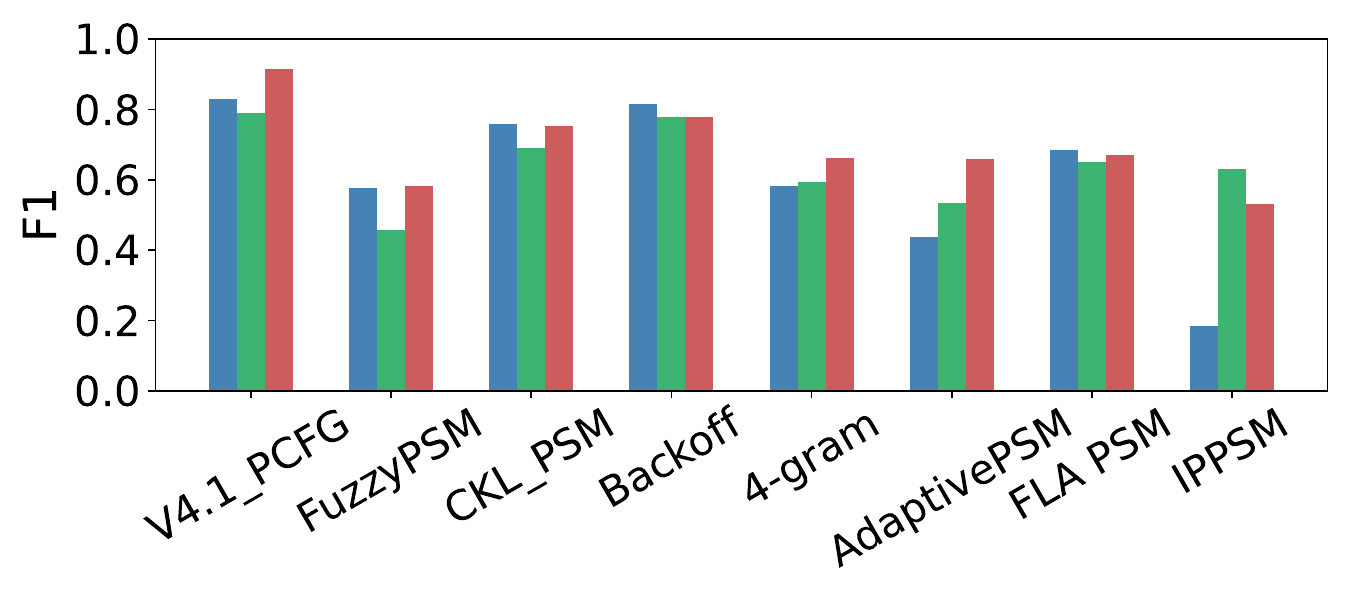}
    }
    \subfigure[F1 (\texttt{XATO})]{
        \includegraphics[width=0.45\linewidth]{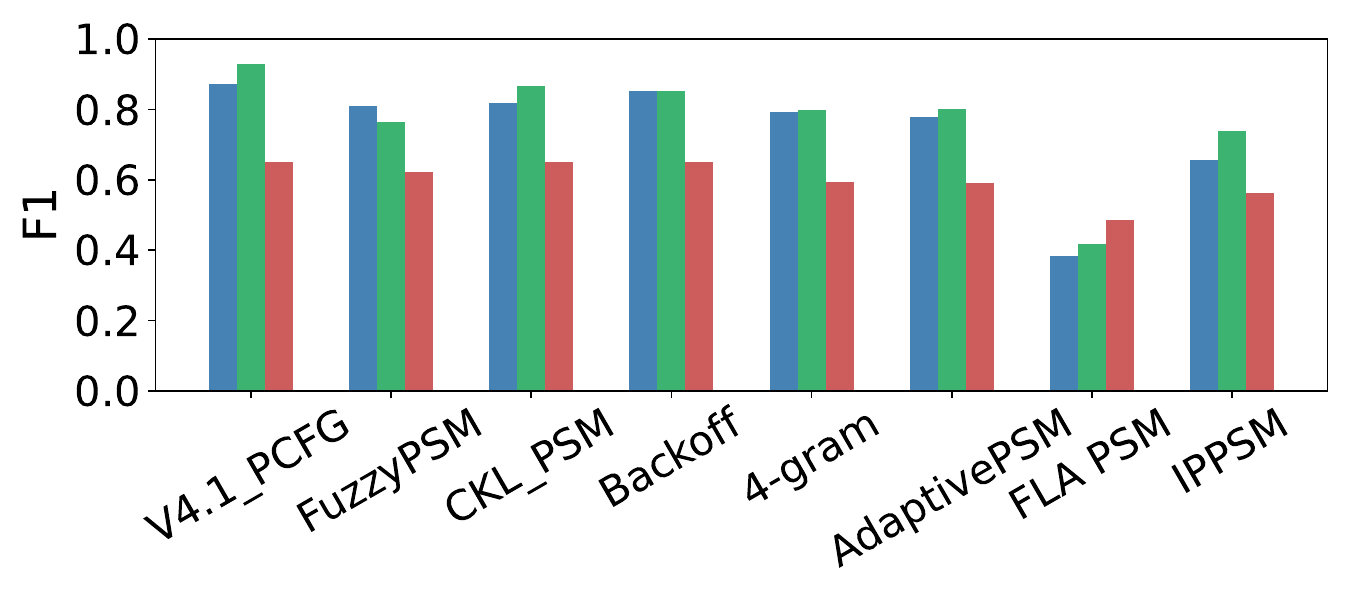}
    }
    \subfigure[Precision (\texttt{178})]{
	\includegraphics[width=0.45\linewidth]{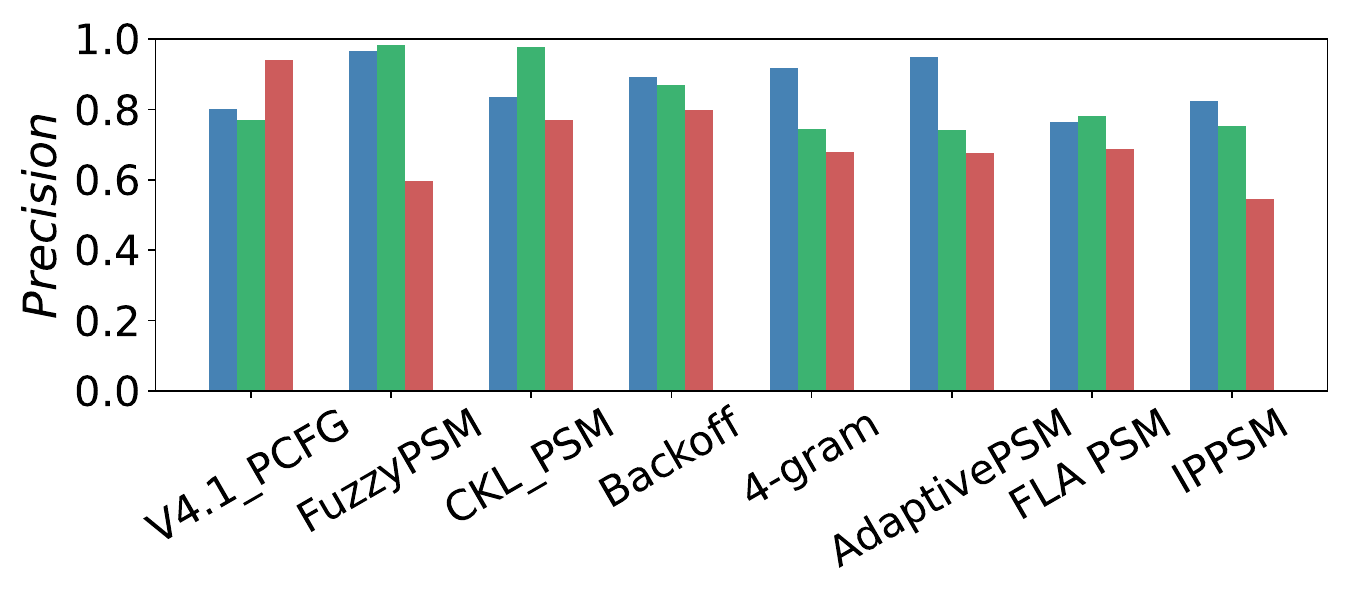}
    }
    \subfigure[Precision (\texttt{XATO})]{
	\includegraphics[width=0.45\linewidth]{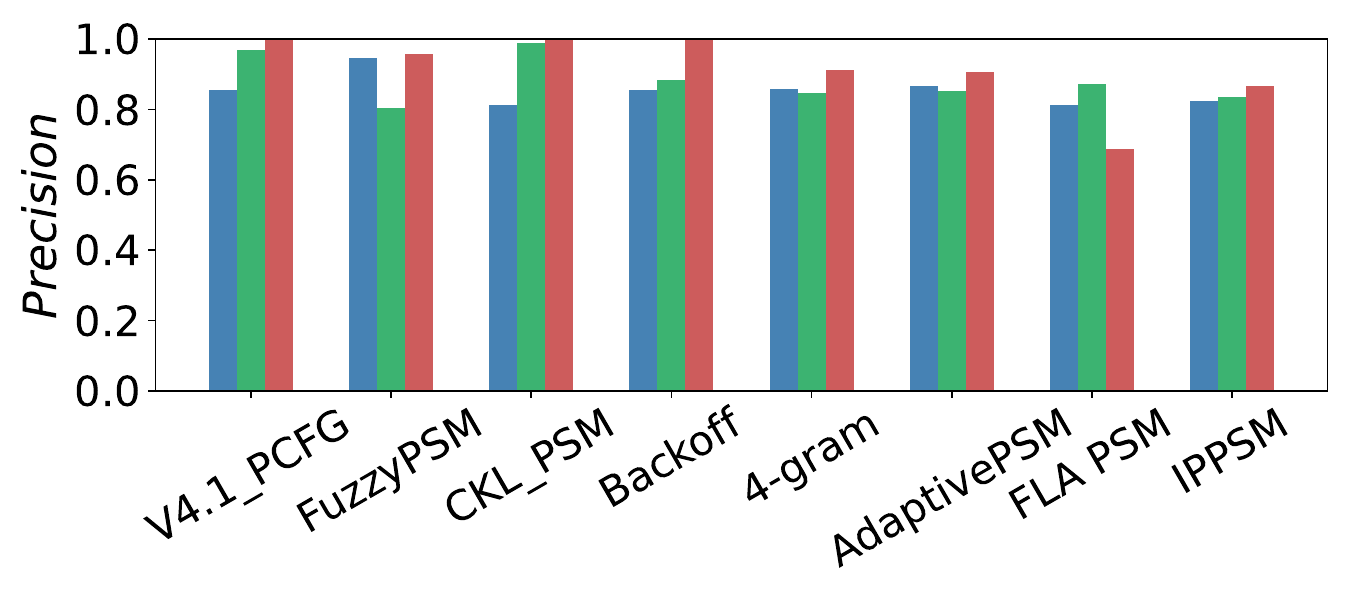}
    }
    \subfigure[Recall (\texttt{178})]{
        \includegraphics[width=0.45\linewidth]{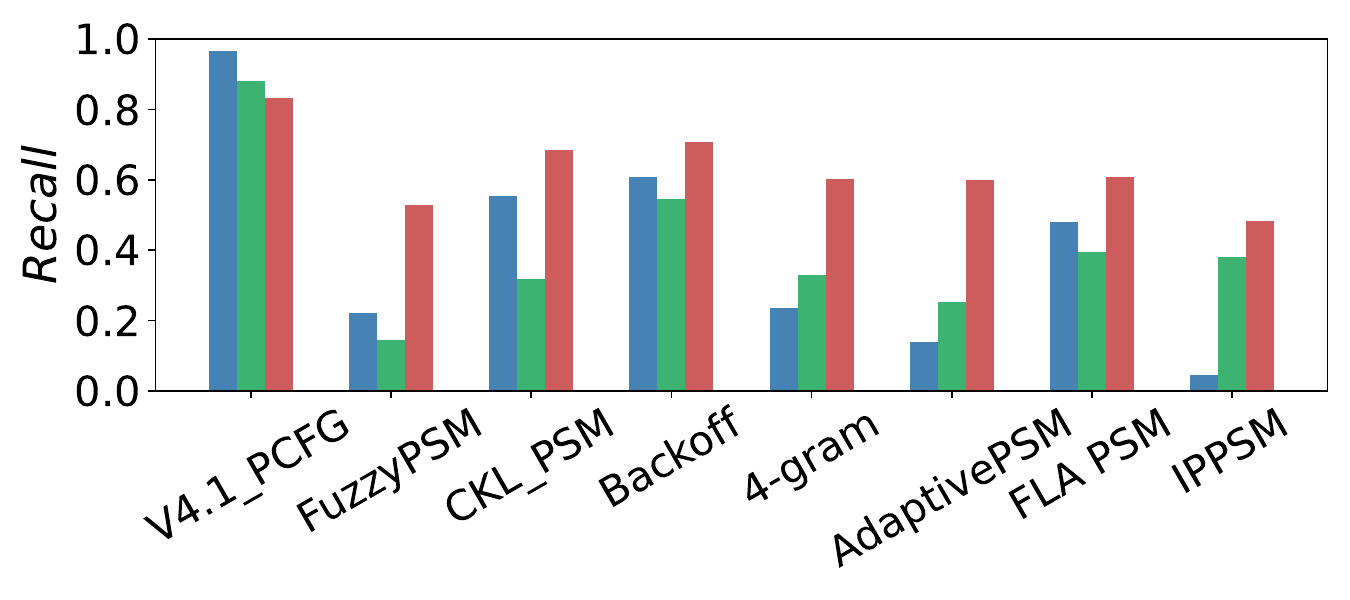}
    }
    \subfigure[Recall (\texttt{XATO})]{
        \includegraphics[width=0.45\linewidth]{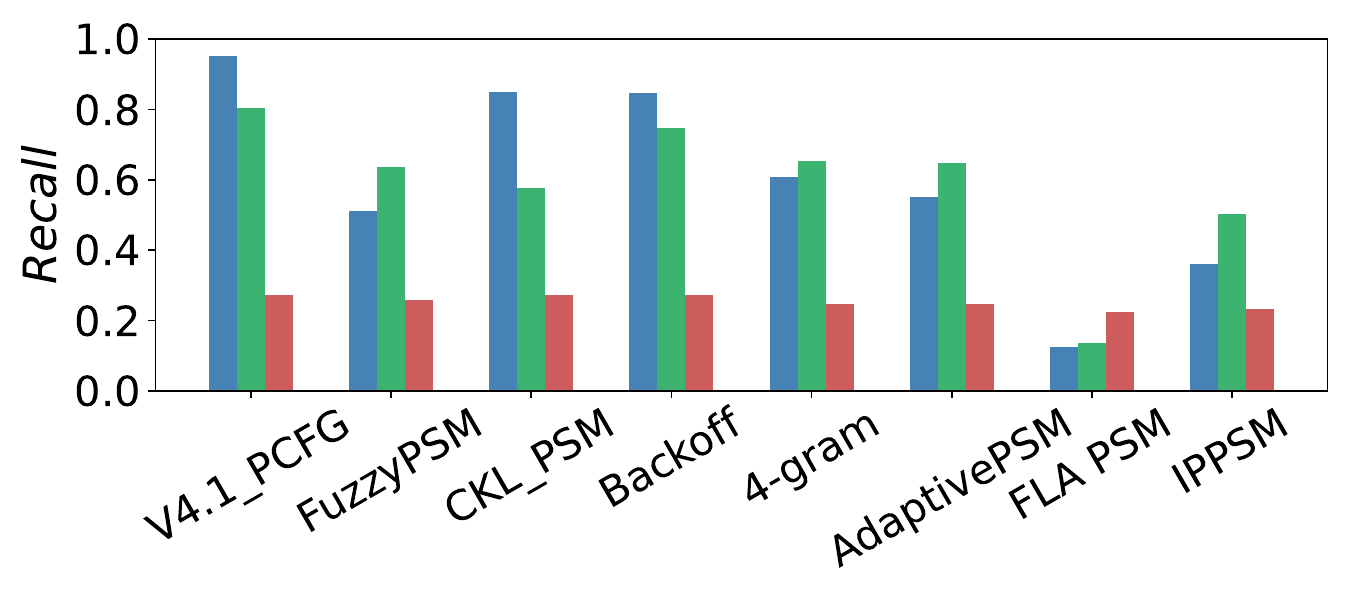}
    }
    \caption{Comparison of the three most effective membership inference attack approaches using optimal parameters.}
    \label{fig:ex_compare_mias}
\end{figure*}

We compare three approaches of threshold-choosing, binary-classifier, and Salems' methods~\cite{DBLP:conf/ndss/Salem0HBF019ML-leaks}. We set the expected precision of 80\% in threshold-choosing methods, choose the jointed features ($J(x|M)$)  and k=10 in Salems' methods, and show the comparison results in Figure~\ref{fig:ex_compare_mias}. Based on this results, we can find that threshold-selection approaches can usually achieve better attack gains than these counter-parts, e.g., Salems' method yields a expremely low recall. 
This could because that Salems' methods lack the information about the target PSM that result in the significant different results on meters. Our threshold-choosing method learns the information from shadow model, and can produce stable results. 
Finally, we compare the average inference time between the three MIA approaches in Table~\ref{table:experiment_performance}, where the inference is calculated as the average inference time across $10^4$ passwords.


\begin{table}[!htb]
\setlength{\abovecaptionskip}{0pt} 
\setlength{\belowcaptionskip}{0pt}
\footnotesize
\renewcommand\tabcolsep{7.8pt} 
\centering
\caption{Overhead of three MIA approaches.} 
\label{table:experiment_performance}
\begin{tabular}{c|c|c}
\toprule
Approaches & Training & Inference\\ \midrule
Probability-threshold-selection     &   35.6 ms   &   2.5 ms   \\ 
Binary classifier    &  $1.7 \times 10^5$ ms    &   271.3 ms   \\ 
Salem's method &   0 ms   &   4.2 ms   \\ 
\bottomrule
\end{tabular}
\end{table} 

\subsection{Principle Stealing Bounds}~\label{app:stealing-upper-bound} 
We initiate to analyze the optimal stealing attacks, wherein attackers can access to all possible passwords in $\pwddict$ and have the ability to query the target probabilistic meters with unlimited number of attempts.
This optimal scenario serves as the upper bound of stealing attacks but is unattainable in the real world due to constraints on computing resources. Given this perfect setting, we quantify the proportion of stolen passwords among attempted guesses ($G$) to analyze the upper bound.    
Given that the optimal sequence of attempted passwords is organized in descending order of probability based on all estimated probabilities from the entirety of password space $\pwddict$ within the target meter, we should find an algorithm to simulate the perfect attempted sequence since it is practically infeasible to compute all estimated probabilities for every passwords from $\pwddict$ modeled by the meter.
We theoretically proof that the top $G$ optimal attempted guesses is equal to the top $G$ password candidates generated by the meter itself. equal to the top password candidates generated by the meter itself.
Formally, we define the top $G$ optimal attempted guesses as $A_{G}(f,\pwddict)$, denoting that they are chosen from $\pwddict$ based on their returned probabilities from $f$. Then, we can deduce that:

\begin{equation}\nonumber
    A_{G}(f,\pwddict) = \{ x_1, x_2, \dots , x_G \},  f(x_1) \ge f(x_2) \ge \dots \ge f(x_G)
\end{equation}

\noindent where $x_i \in \pwddict$.

\noindent By definition~\ref{Definition2} of $C_{G}(f)$ that represents the top $G$ password candidates generated by the meter $f$ in Section~\ref{sec:background}, we can conclude:  

\begin{equation}\nonumber
    A_{G}(f,\pwddict) = C_{G}(f) 
\end{equation}

\begin{figure}
\setlength{\abovecaptionskip}{0pt} 
\setlength{\belowcaptionskip}{0pt}
    \centering
    \includegraphics[width=\linewidth]{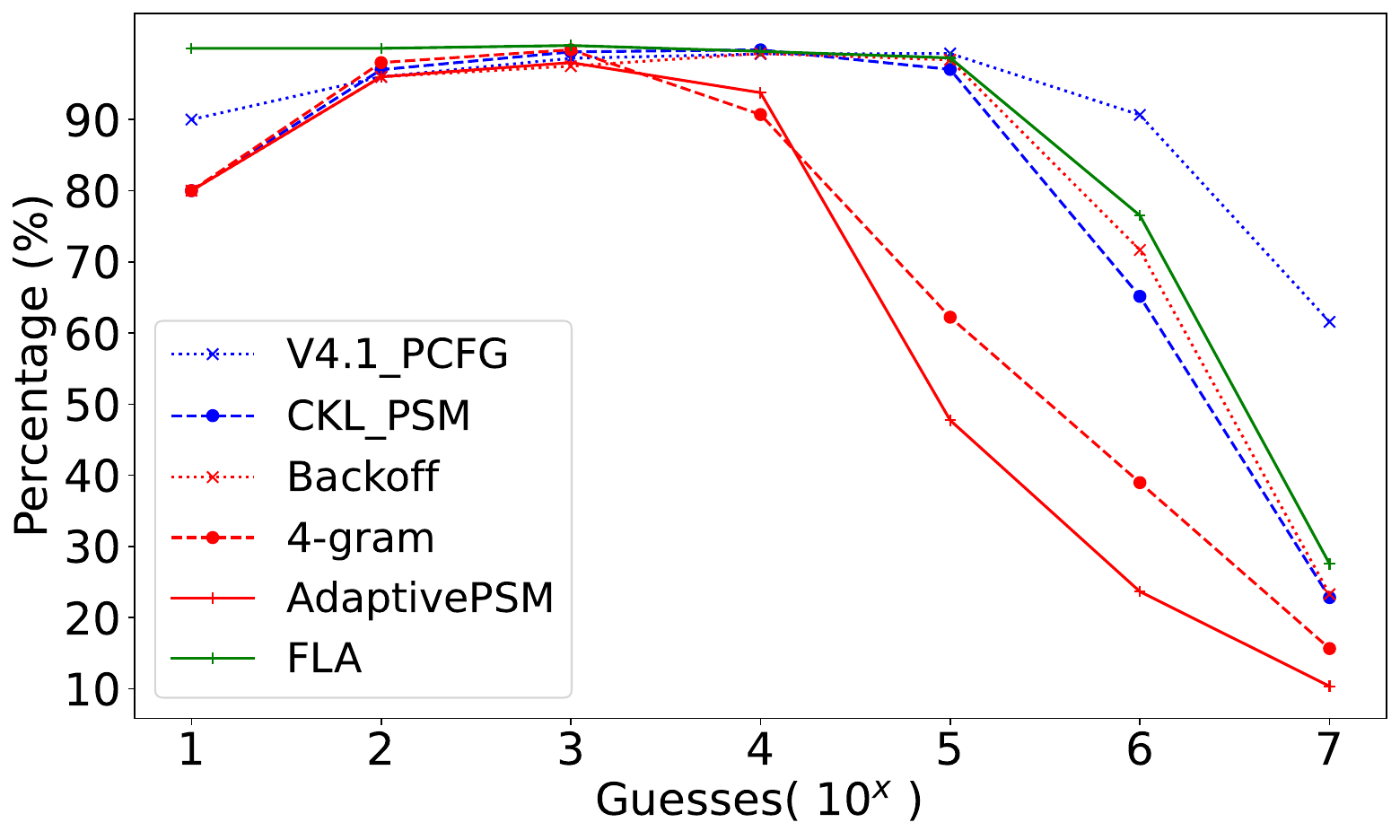}
    \caption{The y-axis shows the maximum percentage of trained passwords among the $10^x$ attempted guesses plotted on the x-axis.}
    \label{fig:ideal-stealing}
\end{figure}

We employ the target model (trained on \texttt{Rockyou}) to generate the top $G=10^7$ password candidates in descending order, and then observe the proportion of trained passwords, where the generation size $G=10^7$ based on the same magnitude of a general password dataset.  
We show the upper bound analysis in Figure~\ref{fig:ideal-stealing}, from which we can find that, when attempting approximately $10^4$ guesses, at most 90\% of trained passwords across these meters can be identified. 
Consistent with conclusions drawn from MIAs, AdaptivePSM exhibits a lower stolen rate compared to its counterparts. 
Besides, Figure~\ref{fig:ideal-stealing} can also quantify the max number of stolen passwords when setting a precision to identify member passwords. The Y-axis also functions as the precision ($\frac{|D_{train} \cap A_G(f,\pwddict)|}{G}$). On the other hand, when we expect at least 90\% precision in stealing trained passwords, we can generally steal a maximum number of around $10^4$ to $10^5$ passwords across these meters.

\section{Recognizing Pattern Passwords}~\label{app:proof}

\noindent\textbf{Identifying name and date patterns.} 
We identify passwords containing name patterns through partial matching with the name dictionary provided in~\footnote{\url{https://github.com/dropbox/zxcvbn}}.
To identify date patterns, we pay attention to length-4, 6 and 8 digits in passwords for their widely usage. We detect date pattern using regular expression matching to formats such as MMYY, MMYYYY, and MMMMYYYY. 
We limit our matching techniques to the year of 1900-2023, month of 1-12, and day of 1-31 to reduce false positives. Furthermore, we note that there might be some false positives, since it is hard to definitely tell apart whether some digit sequences are dates or not, e.g., ``010101'' and ``520520''. These two sequences may be dates, yet they are also likely to be of other semantic meanings (e.g., 520520 sounds like ``I love you I love you'' in Chinese). We refer to the prior literature~\cite{DBLP:conf/uss/usenix19:wangding} that considers those dates when their occurring frequency is larger than a threshold as the valid dates. We assume that user birthdays are randomly distributed in the date patterns.

\begin{table}[]
\setlength{\abovecaptionskip}{0pt}
\setlength{\belowcaptionskip}{0pt}
\caption{Percentage of personal information.}
\label{tab:personal-percentage}   
\renewcommand\tabcolsep{5.9pt}
\footnotesize
\begin{tabular}{ccccccc}
\toprule[\thickline]
\rowcolor{gray!45} Service/Datasets    &   & \begin{tabular}[c]{@{}l@{}}Name\\ percentage\end{tabular} &  & \begin{tabular}[c]{@{}l@{}}Date \\ percentage\end{tabular}  & & \begin{tabular}[c]{@{}l@{}}Phone number\\ percentage\end{tabular} \\
\cmidrule{1-1} \cmidrule{3-3} \cmidrule{5-5} \cmidrule{7-7}
KeePSM           &       &  27.39\%                        &  & 0.01\%  &  & 0\% \\
Zxcvbn           &       &  25.08\%                        &  & 0.64\%   &  & 0.04\% \\   
CUPS PSM &    & 15.14\%                          &  & 18.26\%  &  & 0.04\% \\     
 \cmidrule{1-1} \cmidrule{3-3} \cmidrule{5-5} \cmidrule{7-7}
Popular-PSM &   & 29.61\% &  & 0\%  &  & 0\% \\    
\midrule[\thinline] 
\texttt{Rockyou}          &    &  30.14\%                           & & 0.78\% &  & 1.86\%                            \\
\texttt{XATO}          &    &  9.74\%                           & & 0.76\% & &  0.12\%    \\ 
\texttt{Cit0day}   & & 13.81\%
 &                         & 0.61\% & & 0.07\% \\
\bottomrule[\thickline]
\end{tabular} 
\end{table}

\noindent\textbf{Identifying phone number patterns.} 
We identify the phone numbers of American and English, as they are mainly English-speakers. 
For phone numbers of American, we use the rules of area code (2XX) appending other numbers, i.e., 2XX+XXX-(-)-XXXX with a limit of different numbers in a sequence and an optimal symbol of ``-''. 
For phone numbers of English, we match ten/eleven numbers starting with 07, i.e., 07XXXXXXX(X). Besides, because calls to the UK outside the UK do not need to add 0, only 44, so there is an additional match of nine digits starting with 7, i.e., 7XXXXXXXX(X), and eleven digits starting with 447, i.e., 447XXXXXXX(X).
We empirically find that when searched for the top phone numbers in Google or other similar search engine, some phone numbers can link with a social media like Instagram, posing privacy risks. We do not show the specific phone numbers due to ethical considerations. 

\end{document}